\RequirePackage{fix-cm}
\documentclass[twocolumn,epjc3]{svjour3}  
\smartqed  
\RequirePackage{graphicx}
\journalname{Eur. Phys. J. B}
\begin{document}

\title{Quantum-tunneling transitions and exact statistical mechanics of bistable systems with parametrized Dikand\'e-Kofan\'e double-well potentials
}


\author{F. Naha Nzoupe\thanksref{e1,addr1}
        \and
        Alain M. Dikand\'e\thanksref{e2,addr1} 
        \and S. E. Mkam Tchouobiap\thanksref{e3,addr1}
}

\thankstext{e1}{e-mail: fernand.naha.nz@gmail.com}
\thankstext{e2}{e-mail: dikande.alain@ubuea.cm (Corresponding author)}
\thankstext{e3}{e-mail: esmkam@yahoo.com }

\institute{Laboratory of Research on Advanced Materials and Nonlinear Science (LaRAMaNS), Department of Physics, Faculty of Science, University of Buea P.O Box 63 Buea, Cameroon \label{addr1}
}

\date{Received: date / Accepted: date}

\abstractdc{We consider a one-dimensional system of interacting particles (which can be atoms, molecules, ions, etc.), in which particles are subjected to a bistable potential the double-well shape of which is tunable via a shape deformability parameter. Our objective is to examine the impact of shape deformability on the order of transition in quantum tunneling in the bistable system, and on the possible existence of exact solutions to the transfer-integral operator associated with the partition function of the system. The bistable potential is represented by a class composed of three families of parametrized double-well potentials, whose minima and barrier height can be tuned distinctly. It is found that the extra degree of freedom, introduced by the shape deformability parameter, favors a first-order transition in quantum tunneling, in addition to the second-order transition predicted with the $\phi^4$ model. This first-order transition in quantum tunneling, which is consistent with Chudnovsky's conjecture of the influence of the shape of the potential barrier on the order of thermally-assisted transitions in bistable systems, is shown to occur at a critical value of the shape-deformability parameter which is the same for the three families of parametrized double-well potentials. Concerning the statistical mechanics of the system, the associate partition function is mapped onto a spectral problem by means of the transfer-integral formalism. The condition that the partition function can be exactly integrable, is determined by a criterion enabling exact eigenvalues and eigenfunctions for the transfer-integral operator. Analytical expressions of some of these exact eigenvalues and eigenfunctions are given, and the corresponding ground-state wavefunctions are used to compute the probability density which is relevant for calculations of thermodynamic quantities such as the correlation functions and the correlation lengths.}

\maketitle

\section{Introduction}
\label{intro}
In low-dimensional systems, structural phase transitions are triggered by symmetry breakings that induce phase instabilities under specific conditions \cite{b1,b2,dis1,dis2,dis3,dis4,dik1,dik2}. In general these instabilities are governed by changes in equilibrium properties of the systems, and in the displacive regime \cite{dis1,dis2,dis3,dis4} they can be accompanied by defects in their topological structures including disclinations, discommensurations, domain walls, large-amplitude charge-density waves (phasons) and instantons to name these few ones. Characteristic properties of these structural defects, which are usually referred to as solitons or solitary waves \cite{1,2}, have attracted a great deal of attention over the past years given their widespread applications in a broad range of physical contexts \cite{1,2}. \par In the specific context of condensed matter Physics, the role of instantons in tunneling processes observed in some of these systems at low temperatures is now well established \cite{3,4,10,4a}.  As a matter of fact, in multi-state systems the transitions between two states separated by an energy barrier usually occur either in the classical regime, where they are driven by thermal activations, or in the quantum regime driven by tunneling processes. At high temperatures, classical thermal activations govern the transition which occurs as a hopping over the barrier. However, as the temperature approaches zero, quantum fluctuations gain in importance and the transition is driven by quantum tunnelings through the barrier. However in the later case, the system dynamics can be described by classical configurations which are either instantons or vacuum, that dominate the thermal rate \cite{3,4,10,4a,a}. As the temperature increases, thermally-induced crossovers become more and more relevant and at some critical temperature a phase transition in quantum tunneling can take place \cite{10,4a,a}. The phenomenon of phase transition in quantum tunneling has been extensively investigated over the past years \cite{10,4a,a}, thus it is known that some physical systems can exhibit not only a smooth second-order transition in quantum tunneling at a critical temperature $T_{0}^{(2)}$, but a transition of the first order can also occur \cite{10,5,6,7,8,9,11}. Instructively, this first-order transition in quantum tunneling is closely related to the variation of the instanton action with respect to the instanton energy \cite{10,5,6,7,8,9,11}.
\par
Another relevant issue in the study of phase transitions in the presence of topological defects (i.e. kink solitons), is the statistical mechanics in the relevant temperature regions where phase transitions are expected. Addressing this issue \cite{e,12,13,14,15} a phenomenological theory was introduced, that rests on the functional-integral operator formalism and assumes that at low temperature, the statistical mechanics of one-dimensional systems with kinks and phonons can be mapped onto an eigenvalue problem associated with the transfer-integral operator. For classic models such as the sine-Gordon and $\phi^4$ models, the eigenvalue problem of the transfer-integral operator does not admit exact solutions \cite{12,13}, nevertheless their lowest eigenstates, obtained via approximate methods such as the WKB approximation, can permit to construct acceptable statistical mechanical quantities \cite{12}. Since in the theory of critical phenomena phase instabilities are predicted to occur at temperatures where the free energy, and/or its derivatives representing thermodynamic functions, are nonanalytic, an appropriate formulation of the partition function for a given system therefore represents a key prerequisite for an accurate description of phase instabilities that might take place in the system.  \par
While theoretical approaches used in the studies of phase transitions in quantum tunneling, and of the statistical mechanics of one-dimensional systems, are usually well accepted, the mathematical models describing these physical systems are sometimes weak. For instance the sine-Gordon model was introduced to describe the formation of solitons in one-dimensional systems with periodic energy landscapes, and to investigate soliton contributions to the statistical mechanics of these specific systems \cite{1,12}. However, due to its rigid profile characterized by fixed barrier height and extrema positions as well as a constant period of the substrate potential, the sine-Gordon model could not be exploited for contexts where shape profiles of the periodic substrate could vary in response to processes intrinsic to the systems \cite{1}. To overcome the weakness related to its rigid shape profile, a parametrized version of the sine-Gordon model was proposed i.e. the so-called Remoissenet-Peyrard potential \cite{rp1,rp2,rp3}. On the other hand the $\phi^4$ model was introduced by Landau in his mean-field theory of second-order phase transitions \cite{14,15}. Like the sine-Gordon model, the rigid shape of this double-well model (fixed barrier height and minima positions) restricts its scope to a very limited number of real bistable systems. This limitation motivated the proposal of bistable models with deformable double-well (DW) shapes.\par 
Besides the famous double-Morse model \cite{dm}, introduced in the study of hydrogen-bonded systems such as organic ferroelectrics \cite{dm1} or DNA conformational transitions \cite{per}, some bistable models with parametric DW potentials are found in the literature \cite{dm,20,21,22,a2,a3,20}. In a series of papers \cite{24,23,23a}, three different families of parametrized DW potentials were proposed by Dikand\'e and Kofan\'e (hereafter referred to as DK potentials) that had in common the $\phi^4$ model as a specific limit. The three models differ in their respective distinct parametrization, and thus offer three possible different deformable features for the DW energy profile: in one family the barrier height is constant but positions of the potential minima can be varied \cite{24}, in the second family the parametrization favours continuous variation of the barrier height without changing the two degenerate minima \cite{23}, and in the third family both the barrier height and minima of the DW potential can be tuned simultaneously \cite{23a}. Given that the three families tend to the $\phi^4$ model when their common deformability parameter tends to zero, investigating phase transitions in quantum tunneling and the classical statistical mechanics of the three families of models, is a relevant exercise susceptible to unveil phenomena that the $\phi^4$ model cannot account for. For instance the $\phi^4$ model predicts the transition in quantum tunneling to be only of second order \cite{a}, and we mentioned the fact that the statistical mechanics of this model is non exact within the framework of the transfer-integral formalism \cite{e}. \par
Concerning the possibility of a first-order transition in quantum tunneling in bistable systems, this issue was addressed in a relevant work by Chudnovsky \cite{7}, who postulated that in general this would depend on the shape of the potential barrier. Exploiting Chudnovsky's conjecture, Zhou et al \cite{a} examined the problem by considering a parametrized double-well potential with a variable barrier height but fixed positions of potential minima \cite{23}. They concluded that the increase of the barrier height with the deformability parameter $\mu$, is actually the main factor favoring a first-order transition in quantum tunneling in this particular model of bistable system. However in a recent work \cite{na}, we carried out a study on the same problem using a version of the parametrized double-well potential with fixed barrier height and fixed potential minima, but with variable shape of the barrier top. We obtained that despite the fixed barrier height and fixed minima positions, a first-order transition in quantum tunneling could still be observed in this other model of bistable system. \par Motivated by the two recent studies \cite{23,na}, we have undertaken to investigate conditions for the occurrence of a first-order transition in quantum tunneling for the three distinct members of the so-called {\bf Dikand\'e-Kofan\'e (DK)} potentials \cite{24,23,23a}. The present work will address this issue. We shall also investigate values of the shape deformability paramater, for which exact wavefunctions and eigenvalues of the transfer-integral operator associated with the three distinct parametrized DK potentials, can be found as done recently with the new member of this family introduced in ref. \cite{na}, characterized by a fixed barrier height and fixed minima positions, but a tunable flatness of the barrier top. 
\section{\label{ONE} Three families of parametric double-well potentials}
One-dimensional bistable systems can be represented as a linear chain of particles (atoms, molecules or ions) interacting via two-body forces, with the individual particles lying in one-body potentials with a DW profile. In some physical contexts, shapes of the DW potential can be tuned. For instance particles in the chain can interact with their neighbours via bonds whose lengths are constantly varying (e.g. bond stretching or compression processes), due to variations of pressure, isotopic substitutions or chemical reactions that inevitably affect equilibrium properties of particles along the chain. \par
\begin{figure*}\centering
\begin{minipage}[c]{0.33\textwidth}
\includegraphics[width=2.15in,height=2.in]{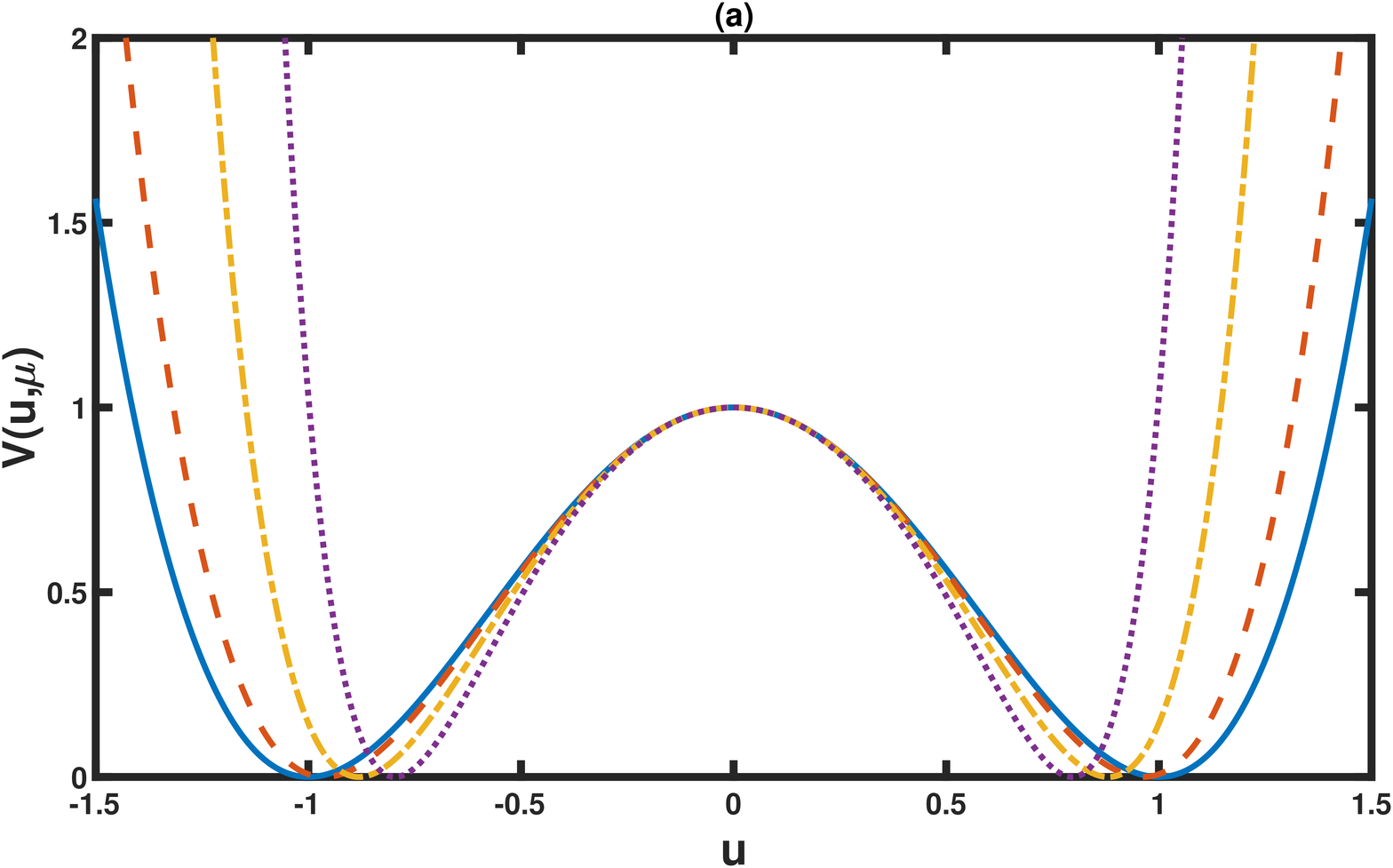}
\end{minipage}%
\begin{minipage}[c]{0.33\textwidth}
\includegraphics[width=2.15in,height=2.in]{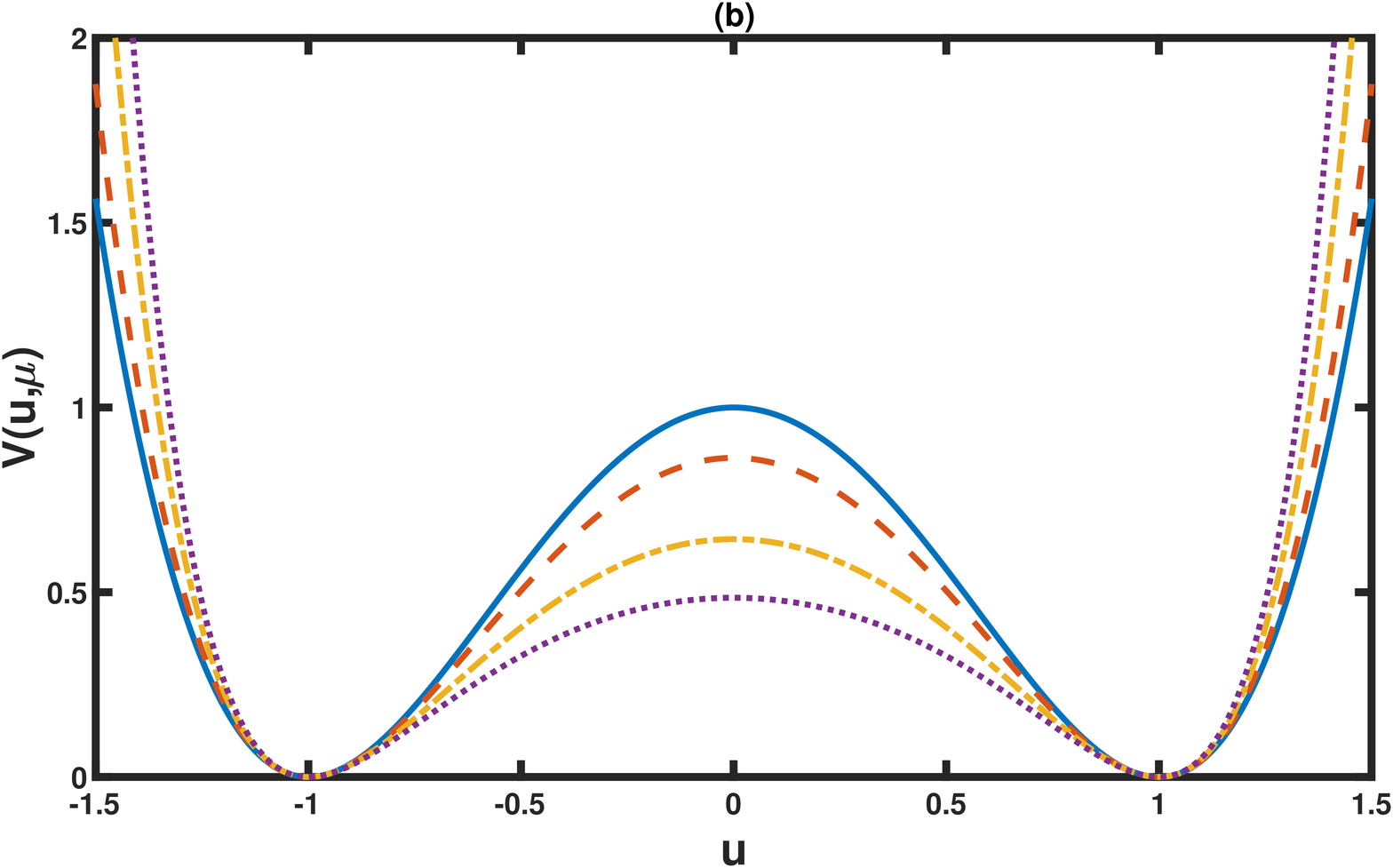}
\end{minipage}%
\begin{minipage}[c]{0.33\textwidth}
\includegraphics[width=2.15in,height=1.8in]{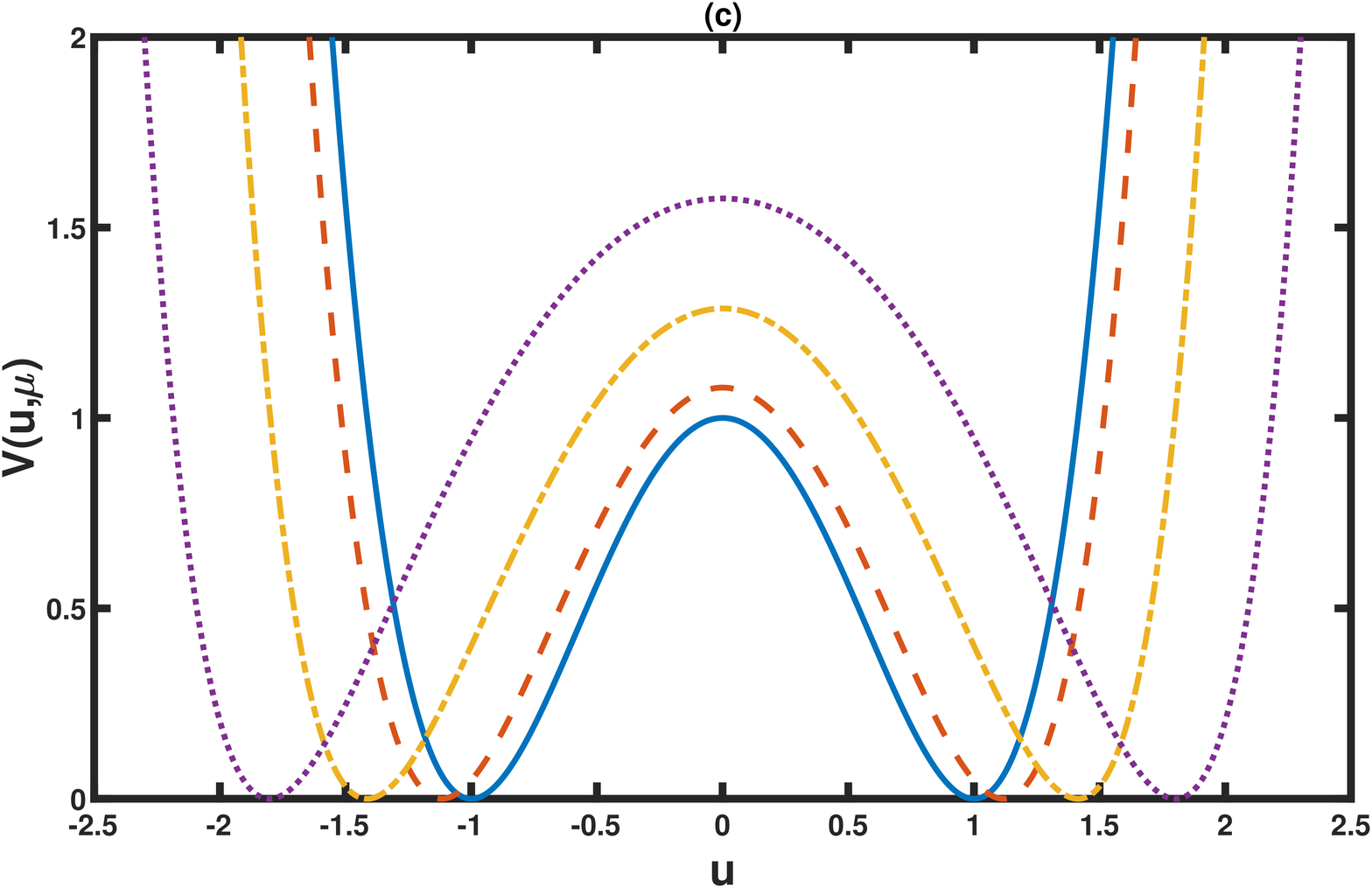}
\end{minipage}
\caption{(Colour online) Sketch of the three families of parametrized DW potentials: (a) Fixed barrier height but variable minima positions, (b) variable barrier height but fixed minima, and (c) simultaneously varying barrier height and positions of generate minima. Values of the deformability parameter $\mu$ are: $\mu\rightarrow 0$ (Solid line, corresponding to the $\phi^4$ potential),  $\mu=0.5$ (Dashed line), $\mu=1.0$ (Dot-dashed line),  $\mu=2.0$ (Dotted line).} 
\label{fig1}
\end{figure*}

 In the present study we are interested in three specific families of DW potentials with deformable shape profiles \cite{24,23,23a}, which can be expressed most generally as \cite{23a}:
\begin{equation}
V(u,\mu) = a(\mu)\left( \frac{\sinh^{2}(\alpha(\mu) u)}{\mu^{2}} - 1\right)^{2}, \hskip 0.3truecm  \mu \neq 0,  
\label{e1}
\end{equation}
where $a(\mu)>0$ and $\alpha(\mu)$ are real functions of the shape deformability parameter $\mu$. $V(u,\mu)$ is a class of parametric DW potentials whose shapes can be tuned differently, but which admit the $\phi^{4}$ potential \cite{12} as a common asymptotic limit. The first family of parametrized DW potentials associated with $V(u,\mu)$, has the following parameters:
\begin{equation}
 \alpha(\mu)=\mu,  \hskip 0.3truecm a(\mu)= a_{0},
\label{e2}
\end{equation}
$a_0=$ constant. This first model represents a DW potential for which positions of the two degenerate minima can  be varied, leaving unchanged the barrier height \cite{24}. Some shape profiles of this first family of DW potentials are sketched in fig. \ref{fig1}-(a). An increase of $\mu$ maintains the potential barrier fixed at $a_0$, whereas positions of the two potential minima; 
\begin{equation}
u_{(1,2)}= \pm \, \frac{sinh^{-1}(\mu)}{\mu}, 
\end{equation}
are continuously shifted near to each other as $\mu$ is varied. As the figure shows this shift induces a narrowing of potential wells, and as a sequel a reduction of the barrier width. A narrowing of potential wells will enhance confinement of particles inside the wells, thus favoring tunneling processes given that the stiffness of the two walls culminating at the barrier top is expected to increase with the confinement. \par    
For the second family, characteristic parameters are given by:
\begin{equation}
 \alpha(\mu)=sinh^{-1}(\mu),  \hskip 0.3truecm a(\mu)=\frac{a_{0}\,\mu^{2}}{\left[\alpha(\mu)\sqrt{1+\mu^2}\right]^2}.
\label{e3}
\end{equation}
This second model corresponds to a DW potential with fixed degenerate minima, but a tunable barrier height \cite{23}. Fig.\ref{fig1}-(b) illustrates profiles of the second model of DW potential, for some values of the deformability parameter $\mu$. The two minima are indeed always fixed at $u=\pm\, 1$, but when $\mu$ is increased the barrier height continuously decreases with the barrier top gradually flattened. \\
The third model has the following parameter values: 
\begin{equation}
\alpha(\mu)=\frac{sinh^{-1}(\mu)}{\sqrt{1+\mu^2}}, \hskip 0.3truecm a(\mu)=\frac{a_{0}\,\mu^{2}}{\left[sinh^{-1}(\mu)\right]^2}.
\label{e4}
\end{equation}
This third model mimics a DW potential whose degenerate minima and barrier height can be simultaneously varied \cite{23a}. In fig.\ref{fig1}-(c) it is seen that an increase of $\mu$ causes the increase of the barrier height, as well as a shift away of positions of the two potential minima. This third model is clearly most prone to tunneling processes when $\mu$ is large, given that by increasing the deformability parameter the barrier height is increased while the distance from the equilibrium positions of a particle inside the wells to the position of the potential barrier, is relatively large. Therefore it will require more and more energy for a particle to move from inside the wells to the barrier top, in order to hope from one well to the other unless it proceeds via tunneling. 
Remark that When $\mu$ tends to zero, the three families of parametrized DW models reduce to the standard $\phi^4$ model i.e. \cite{12}:
\begin{equation}
V(u)=a_{0}\left( u^{2} - 1\right)^{2}.
\end{equation}
In the next section we investigate phase transitions in quantum tunneling mediated by periodic instantons, for the three families of parametrized DW models. We shall sometimes refer to the three families of parametrized DW potentials as DK potentials \cite{24,23,23a}. 

\section{\label{two}Periodons and transitions in quantum tunneling}
Consider the following Euclidean action for one-dimensional quantum models (in dimensionless unity) \cite{10,4a,a}:
\begin{equation}
S = \int d\tau\left(\frac{1}{2}\left(\frac{du}{d\tau}\right)^{2}+ V(u,\mu)\right),
\label{e5}
\end{equation}
 where $u$ represents a scalar field in one time and zero space dimension. The variable $\tau= it$ is the imaginary time, and $V(u,\mu)$ is a nonlinear potential energy of the general form eq. (\ref{e1}). The integral in eq. (\ref{e5}) is taken over the period $\tau_{p}$ of the trajectory in Euclidean space. In statistical mechanics, the period $\tau_{p}$ is related to temperature $T$ through $\tau_{p} = \hbar /(k_{B}T)$, where $k_{B}$ is the Boltzmann constant. For simplicity and without loss of generality, in this section we shall assume $(\hbar, k_{B})\equiv 1$.\\
To an exponential accuracy, the decay rate in the semiclassical limit is given by:
\begin{equation}
\Gamma \sim \exp(-S_{min}),
\label{e6}
\end{equation} 
where  $S_{min}$ is the minimum effective Euclidean action obtained by minimizing eq. (\ref{e5}), along a trajectory $u_{c}(\tau)$ of the classical pseudoparticle. This trajectory satisfies the equation:
\begin{equation}
\left(\frac{du_{c}}{d\tau}\right)^{2}= 2\left( V(u_{c},\mu )- E\right),
\label{e7}
\end{equation}
where the integration constant $E\geq 0$ can be regarded as the energy of the pseudoparticle of mass unity.
When $E=0$, The trajectory $u_{c}(\tau)$ is a regular vacuum instanton described by the single-kink soliton:
\begin{equation}
u_{c}(\tau)=  \frac{1}{\alpha(\mu)}\tanh^{-1}\left[\frac{\mu}{\sqrt{1+\mu^{2}}} \tanh \frac{\tau}{\sqrt{2}d(\mu)}\right],
\label{e8}
\end{equation}
where:
\begin{equation}
d(\mu)= \frac{\mu}{\left[a(\mu)\alpha^{2}(\mu)(1+\mu^{2})\right]^{1/2}} \label{tail}
\end{equation}
is the kink width. However when $E> 0$, using periodic boundary conditions and identifying $\tau_{p}$ as the period of motion, we find that the instanton solution to eq. (\ref{e7}) should now read: 
\begin{equation}
u_{c}(\tau)=  \frac{1}{\alpha(\mu)}\tanh^{-1}\left[C_{1}\cdot sn\left(C_{2}\,\tau,\kappa \right)\right].
\label{e9}
\end{equation}
In this later expression the quantity $sn(\tau,\kappa )$ is one of Jacobi elliptic functions \cite{jac}, its modulus $\kappa$ is given by: 
 \begin{equation}
\kappa = \sqrt{\frac{\left(1-\sqrt{E/a(\mu)}\right)\left[1+\mu^{2}\left(1+\sqrt{E/a(\mu)}\right)\right]}{\left(1+\sqrt{E/a(\mu)}\right)\left[1+\mu^{2}\left(1-\sqrt{E/a(\mu)}\right)\right]}}. 
\label{e10}
\end{equation}
The two parameters $C_{1}$ and $C_{2}$ in the instanton solution eq. (\ref{e9}) are defined as: 
\begin{eqnarray*}
C_{1} &=&\sqrt{\frac{\mu^{2}\left(1-\sqrt{E/a(\mu)}\right)}{1+\mu^{2}\left(1-\sqrt{E/a(\mu)}\right)}},
\label{e11} \\
C_{2} &=& \frac{\alpha(\mu)}{\mu}\sqrt{2a(\mu)\left(1+\sqrt{ \frac{E}{a(\mu)} }\right)\left[1+\mu^{2}\left(1-\sqrt{  \frac{E}{a(\mu)}   }\right)\right]}.
\label{e12}
\end{eqnarray*}
The instanton trajectory (\ref{e9}) possesses real periods $4mK(\kappa)$, with $m$ an integer and $K(\kappa)$ the quarter period defined in terms of the complete elliptic integral of the first kind \cite{jac}. Hence the trajectory eq. (\ref{e9}) is a periodic instanton, with a period $\tau_p$ corresponding to $m = 1$ and given by:
\begin{equation}
\tau_{p} = 4K(\kappa)/C_2.
\label{e13}
\end{equation}
We refer to the periodic instanton eq. (\ref{e9}) as a periodon. From eq. (\ref{e9}) we can see that the periodon exists only in the range $0\leq E \leq a(\mu)$, and that the expression of the vacuum instanton, i.e. the single-kink soliton eq. (\ref{e8}), will be recovered when $E = 0$ (i.e. $\kappa=1$ \cite{jac}). \\
Using the energy-integral equation (\ref{e7}), the classical action $S_{p}$ of the periodon can be computed and is found to be:
\begin{eqnarray*}
S_{p}(E) &=& 2\int^{\tau/4}_{\tau/4} d\tau\left(\frac{1}{2}\left(\frac{du}{d\tau}\right)^{2}+ V(u,\mu)\right) \nonumber \\
&=& \tau_{p}E + W, \label{e14}
\end{eqnarray*}
where;
\begin{eqnarray}
W &=& \frac{2 C_{2}}{(\alpha(\mu) C_{1})^{2}}\left[(C_{1}^{4}-\kappa^{2})\Pi(C_{1},\kappa)+ \kappa^{2}K(\kappa)\right] \nonumber \\
&+&  \frac{2 C_{2}}{(\alpha(\mu))^{2}}(K(\kappa) - E(\kappa)).
\label{e15}
\end{eqnarray}
In this formula $E(\kappa)$ and $\Pi(C_{1},\kappa)$ are complete elliptic integrals of the second and third kinds, respectively. We note that when $E = a(\mu)$ the trajectory reduces to a trivial configuration $u_{c}(\tau)=0$, which is periodic with an arbitrary period. This configuration, which we shall call sphaleron \cite{b}, has an action coinciding with the thermodynamic action $S_{0}$ given by \cite{b}:
\begin{equation}
S_{0} = a(\mu)\tau.
\label{e16}
\end{equation}
 For the sphaleron configuration, the escape rate is the Boltzmann formula representing a pure thermal activation i.e.:
 \begin{equation}
\Gamma_{c} \sim \exp\left(- a(\mu)\tau_{p}\right) = \exp\left(- a(\mu)/T\right).
\label{e17}
\end{equation}
From the above discussions we can conclude that a periodon (i.e. the periodic instanton) interpolates between the sphaleron and the vacuum instanton (i.e. the single-kink instanton). Therefore the actual escape rate should be of the form of eq. (\ref{e6}), where the minimum action $S_{min}(E)$ is given by:
\begin{equation}
S_{min}(E) = min\left\lbrace S_{0}, S_{p}(E) \right\rbrace.
\label{e18}
\end{equation}
 When dealing with periodic problems in classical statistical mechanics, it is known that the derivative of the action with respect to the energy is equivalent to the oscillation time $\tau$ at this energy, where $\tau$ is proportional to the inverse temperature $T$. Hence, with the action corresponding to the motion in the potentials we find:
 \begin{equation}
\tau_{p}(E) = \frac{1}{T}= \frac{dS_{p}(E)}{dE}, \hskip 0.3truecm  \frac{dS_{0}}{d\tau_{p}} = a(\mu).
\label{e19}
\end{equation}
Taking eq. (\ref{e19}) together with eqs. (\ref{e13}) and (\ref{e14}), we can determine the influence of the shape deformability parameter on the temperature dependence of $S_{min}$, using the dependence of the period of periodon on the energy $E$. In the following analysis these equations will enable us to point out the existence of a critical value of the shape deformability parameter, for which a first-order transition from quantum to thermal regimes can occur for the three families of DKDW models.\par
 Chudnovsky \cite{7} postulated two criteria for the occurrence of transitions in quantum tunneling. The first states that the transition is of first order if $\tau_{p}(E)$ decreases to a minimum and then rises again, when $E$ increases from the potential bottom (i.e. minima) to the barrier top. On the other hand, if the instanton period $\tau_{p}(E)$ is monotonically decreasing when $E$ increases, the transition is of second order.
The second criterion states that if at some critical temperature the first derivative of  $S_{min}(T)$ is discontinuous, and an abrupt change is observed in the temperature dependence of the action, then the transition in quantum tunneling is a first-order transition in temperature. \par Chudnovsky's first criterion is equivalent to solving the equation $d\tau_{p}(E)/dE=0$, which is tractable when seeking nontrivial solutions and considering the temperature dependence of the period given by (\ref{e13}). In a numerical procedure we have chosen $a_{0}= 1/2$, and computed the energy $E_{1}$ of the minimum $\tau_{p}(E)$ by varying the deformability parameter $\mu$ from the value $\mu=3$ downwards. Results for the three models are listed in Tables \ref{table1}, \ref{table2} and \ref{table3}.
\begin{table}[ht]
\caption{Critical values of $E_{1}$ for model 1 (fixed barrier height but variable minima)} 
\centering 
\renewcommand{\arraystretch}{1.2}
\begin{tabular}{c c c c c c c c c c} 
\hline \hline 
 $\mu^{2}$ & & & Energy $E_{1}$ at  & & & $E_{0}= a_{0}$\\ 
  & & & minimum of $\tau_{p}(E)$  & & & (barrier height)\\  
\hline 
 9 & & & 0.150 & & & 0.5\\ 
 4 & & & 0.228 & & & 0.5\\ 
 2 & & & 0.382 & & & 0.5\\ 
 1.6 & & & 0.469 & & & 0.5\\ 
 1.501 & & & 0.499 & & & 0.5\\ [1.ex] 
\hline 
\end{tabular}
\label{table1} 
\end{table}

\begin{table*}[ht]
\caption{Critical values of $E_{1}$ for model 2 (fixed minima but variable barrier height)} 
\centering 
\renewcommand{\arraystretch}{1.2}
\begin{tabular}{c c c c c c c c c c} 
\hline \hline 
 $\mu^{2}$ & & & Energy $E_{1}$ at  & & & $E_{0}= a_{0}\mu^{2}\left[\alpha(\mu)\sqrt{1+\mu^{2}}\right]^{-2}$\\ 
  & & & minimum of $\tau_{p}(E)$  & & & (barrier height)\\  
\hline 
 9 & & & 0.041 & & & 0.136\\ 
 4 & & & 0.087 & & & 0.192\\ 
 2 & & & 0.194 & & & 0.254\\ 
 1.6 & & & 0.258 & & & 0.275\\ 
 1.501 & & & 0.281 & & & 0.282\\ [1.ex] 
\hline 
\end{tabular}
\label{table2} 
\end{table*}

\begin{table*}[ht]
\caption{Critical values of $E_{1}$ for model 3 (variable barrier height and minima)} 
\centering 
\renewcommand{\arraystretch}{1.2}
\begin{tabular}{c c c c c c c c c c} 
\hline \hline 
 $\mu^{2}$ & & & Energy $E_{1}$ at  & & & $E_{0}= a_{0}\mu^{2}\left[sinh^{-1}\mu\right]^{-2} $\\ 
  & & & minimum of $\tau_{p}(E)$  & & & (barrier height)\\  
\hline 
 9 & & & 0.409 & & & 1.361\\ 
 4 & & & 0.437 & & & 0.960\\ 
 2 & & & 0.581 & & & 0.761\\ 
 1.6 & & & 0.672 & & & 0.716\\ 
 1.501 & & & 0.704 & & & 0.705\\ [1.ex] 
\hline 
\end{tabular}
\label{table3} 
\end{table*}
The emerging trend from the three tables is $E_{1}$ tending to the maximum energy $E_{0}$ of the pseudoparticle energy, when $\mu$ approaches $\sqrt{3/2}$. Hence $\mu_{c}=\sqrt{3/2}$ is a critical value of the shape deformability parameter, values of $\mu$ smaller than $\mu_c$ will correspond to forbidden energy configurations (i.e. nonanalytical values of $E$). In Chudnovsky picture a first-order quantum-to-classical transition will occur when $\mu > \mu_c$. When $\mu \approx \mu_{c}$ we obtain 
$E_{1}\approx E_{0}$. Moreover $E_{1}\rightarrow 0$ as $\mu$ is increased above $\mu_{c}$, so \cite{c} increasing the shape deformability parameter above the critical value $\mu_{c}$ results in a sharper first-order transition in quantum tunneling.\par
\begin{figure*}\centering
\begin{minipage}[c]{0.52\textwidth} 
\includegraphics[width=3.in,height=2.5in]{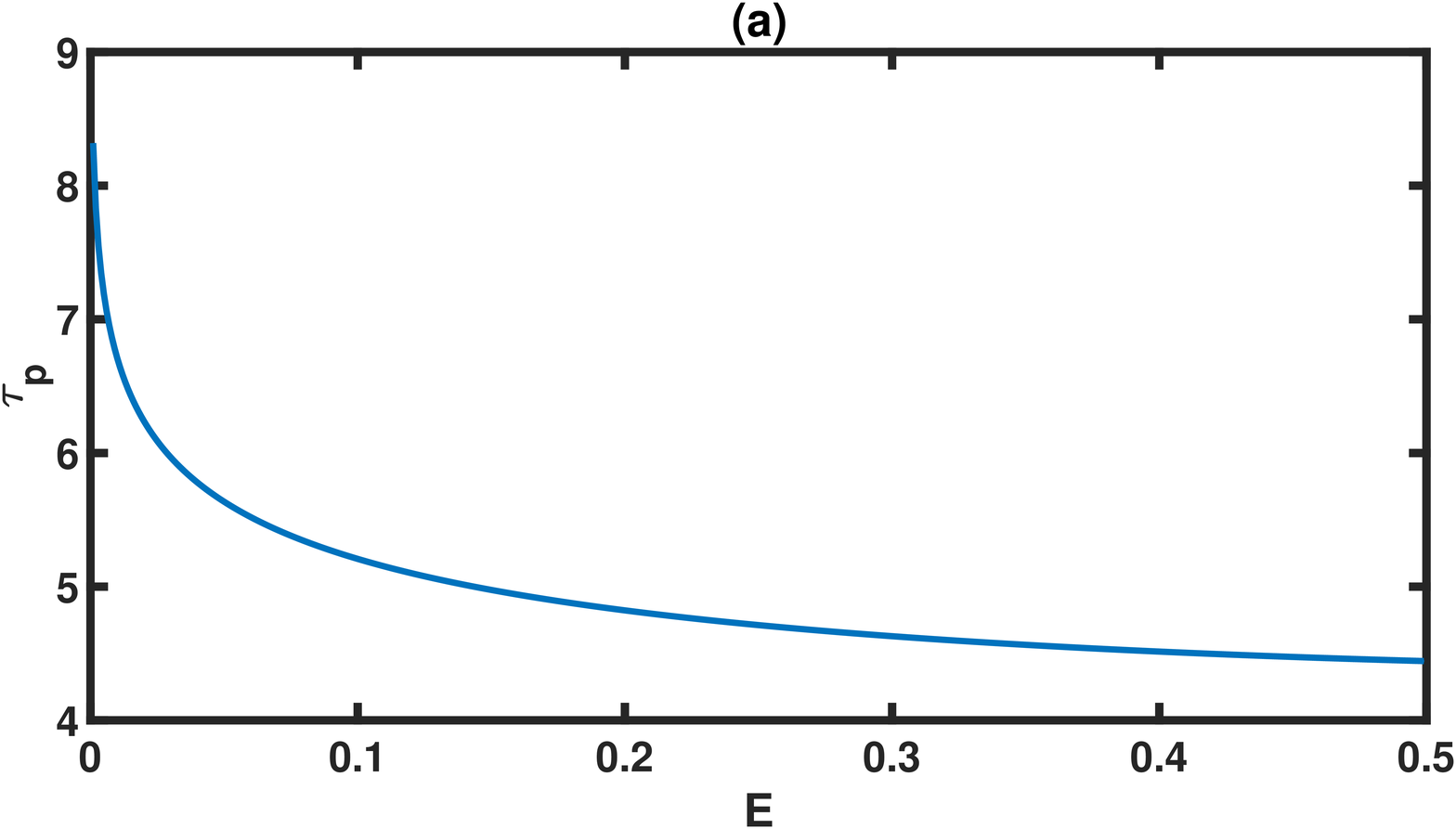}
\end{minipage}%
\begin{minipage}[c]{0.52\textwidth} 
\includegraphics[width=3.in,height=2.5in]{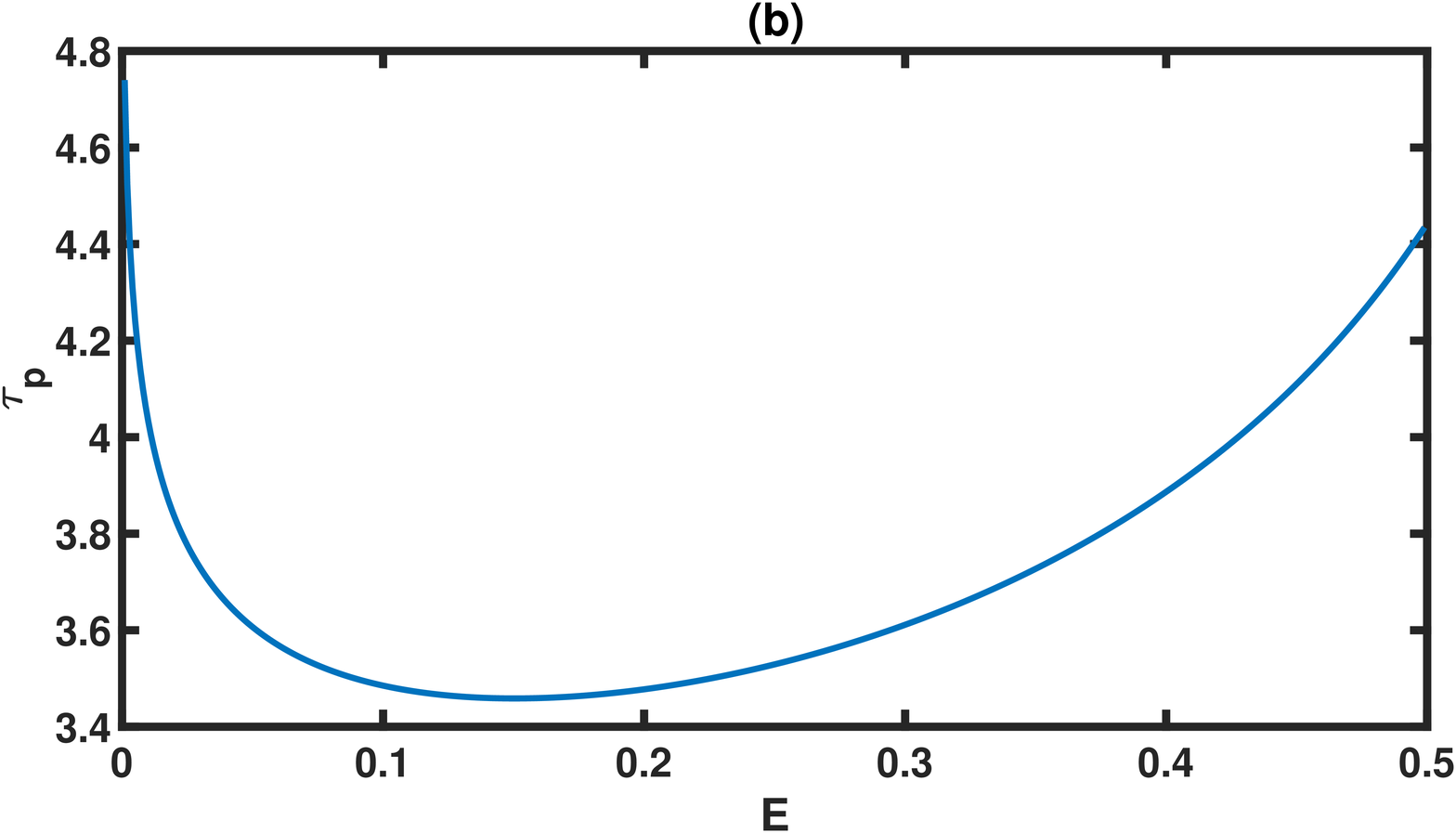}
\end{minipage}
\caption{(Colour online) Plot of the instanton period $\tau_p$ as a function of the energy $E$, for model 1. (a) $\mu=1$, and (b) $\mu=3$. Here $a_{0} = 0.5$.} 
\label{fig2}
\end{figure*}
\begin{figure*}\centering
\begin{minipage}[c]{0.52\textwidth}  
\includegraphics[width=3.in,height=2.5in]{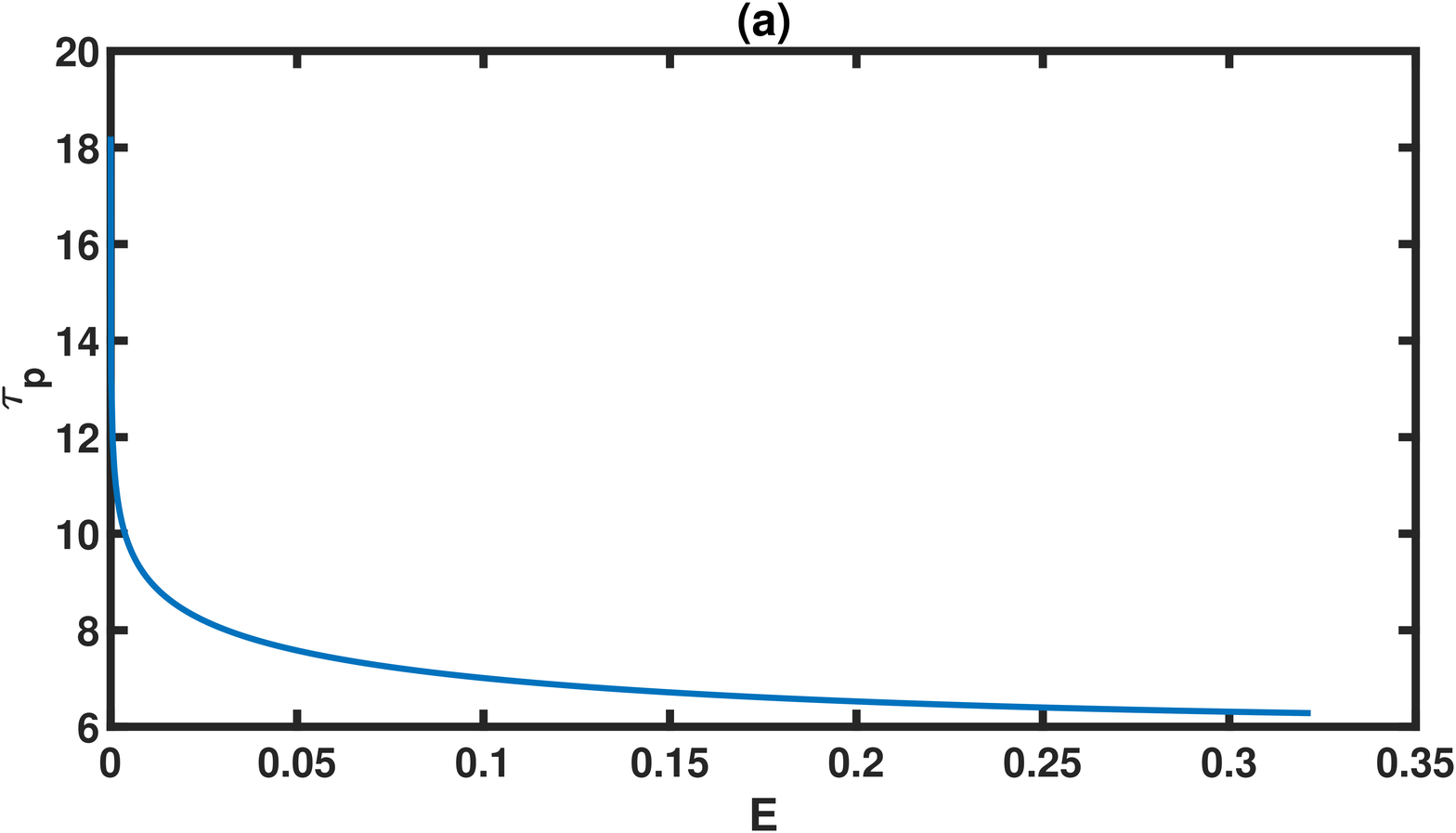}
\end{minipage}%
\begin{minipage}[c]{0.52\textwidth} 
\includegraphics[width=3.in,height=2.5in]{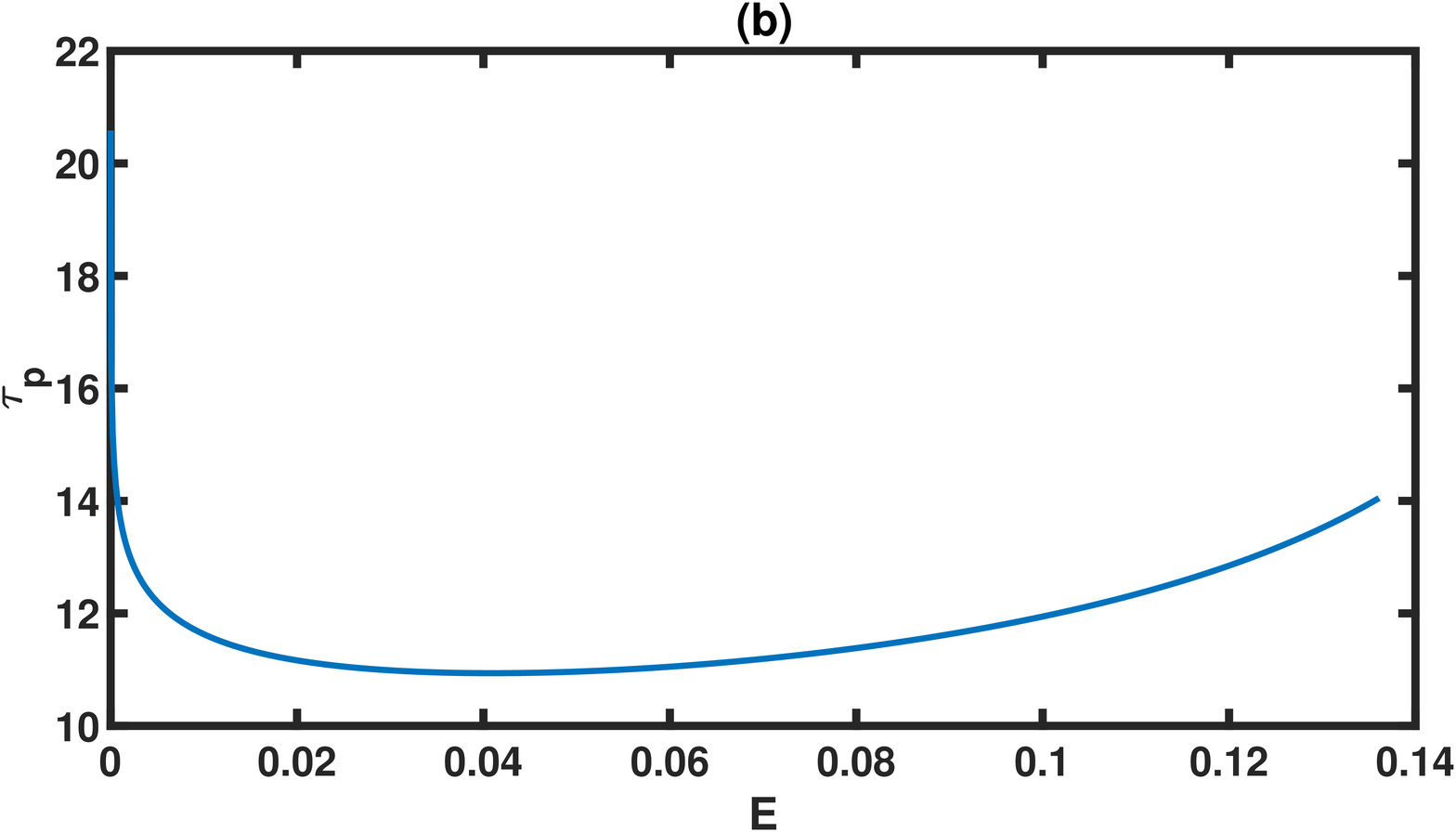}
\end{minipage}
\caption{(Colour online) Plot of the instanton period $\tau_p$ as a function of the energy $E$, for model 2. (a) $\mu=1$, and (b) $\mu=3$. Here $a_{0} = 0.5$.} 
\label{fig3}
\end{figure*}
\begin{figure*}\centering
\begin{minipage}[c]{0.52\textwidth} 
\includegraphics[width=3.in,height=2.5in]{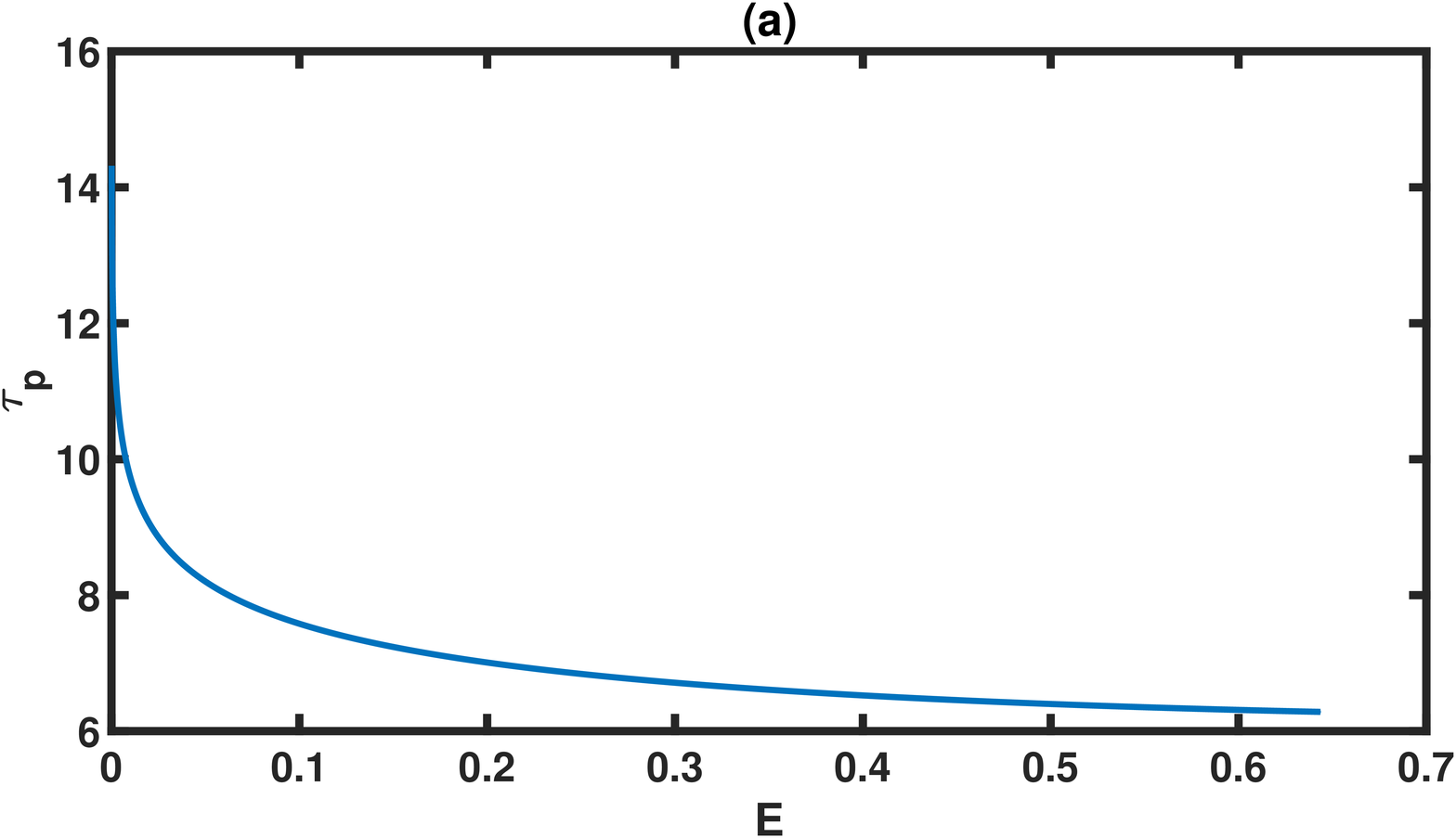}
\end{minipage}%
\begin{minipage}[c]{0.52\textwidth} 
\includegraphics[width=3.in,height=2.5in]{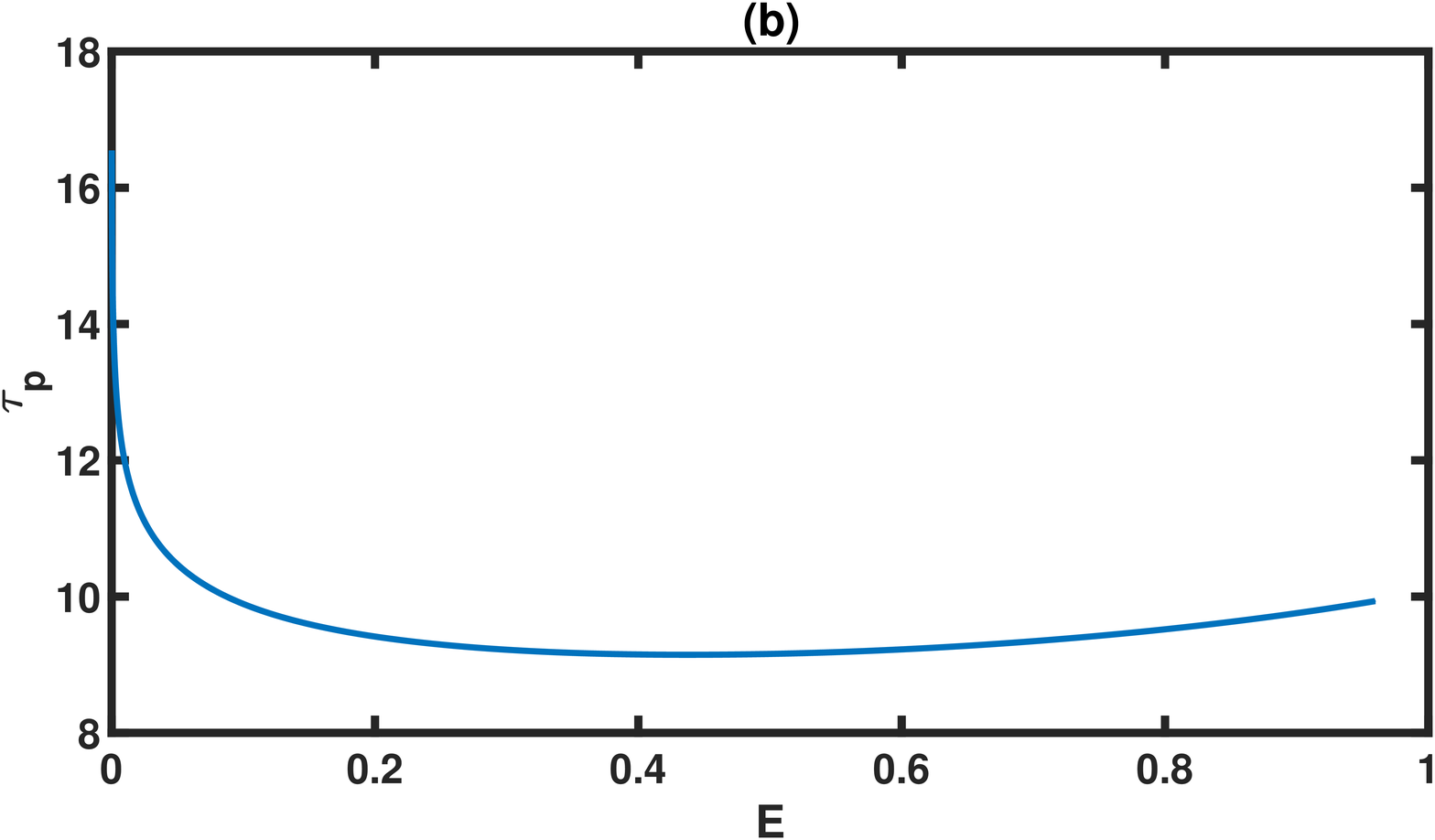}
\end{minipage}
\caption{(Colour online) Plot of the instanton period $\tau_p$ as a function of the energy $E$, for model 3. (a) $\mu=1$, and (b) $\mu=3$. Here $a_{0} = 0.5$.} 
\label{fig4}
\end{figure*}

Figs. \ref{fig2}, \ref{fig3} and \ref{fig4} show the energy dependence of the period $\tau_{p}$ of the periodon for the three models, considering two values of the shape deformability parameter $\mu$ respectively below and above the critical value $\mu_{c}$.
For the three models the figures suggest a monotonic decrease of the period with increasing energy for $\mu = 1$ (a value lower than $\mu_{c}$). However, for $\mu = 3$, we notice an increase of the period after a critical value of the energy $E$. This increase reflects favorable conditions for a first-order transition in quantum tunneling.
\begin{figure*}\centering
\begin{minipage}[c]{0.52\textwidth}
\includegraphics[width=3.in,height=2.5in]{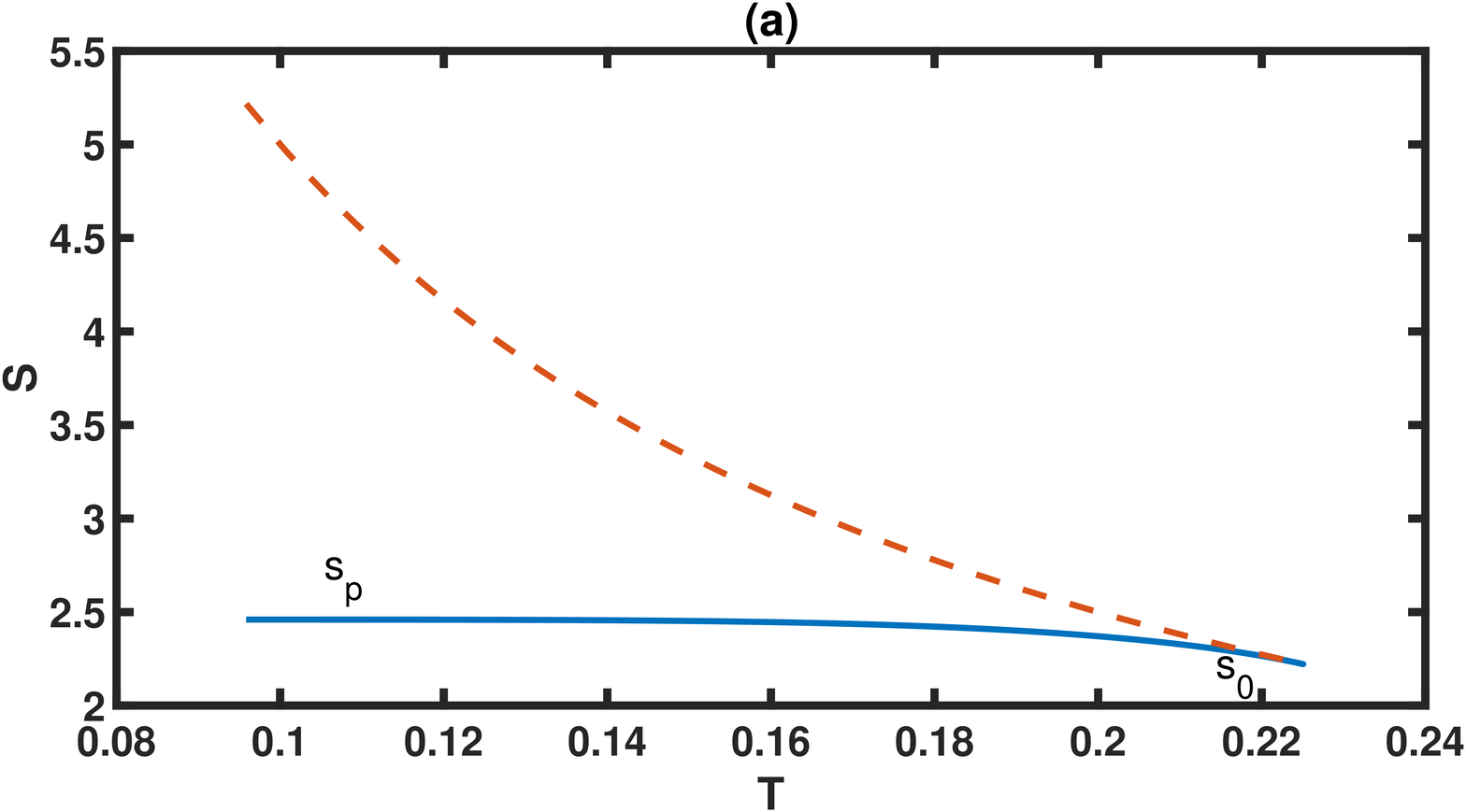}
\end{minipage}%
\begin{minipage}[c]{0.52\textwidth}
\includegraphics[width=3.in,height=2.5in]{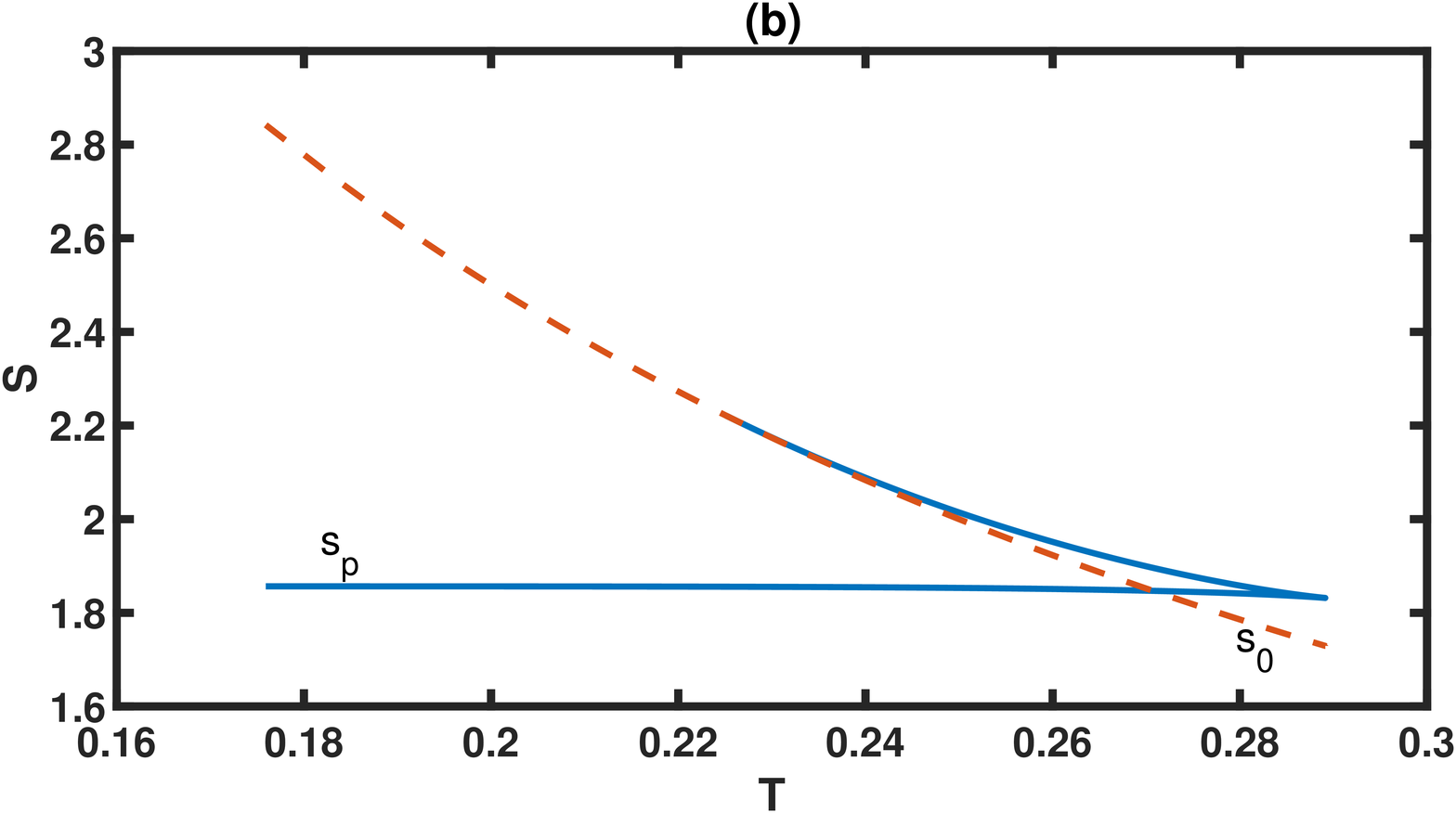}
\end{minipage}
\caption{(Colour online) The action versus temperature for model 1, the dashed line corresponds to the thermodynamic action and the solid line to the periodon action.  (a) $\mu=1$: signature of second-order transition from quantum to thermal regimes,  (b) $\mu=3$: signature of first-order transition from quantum to thermal regimes. Here $a_0 = 0.5$.} 
\label{fig5}
\end{figure*}
\begin{figure*}\centering
\begin{minipage}[c]{0.52\textwidth} 
\includegraphics[width=3.in,height=2.5in]{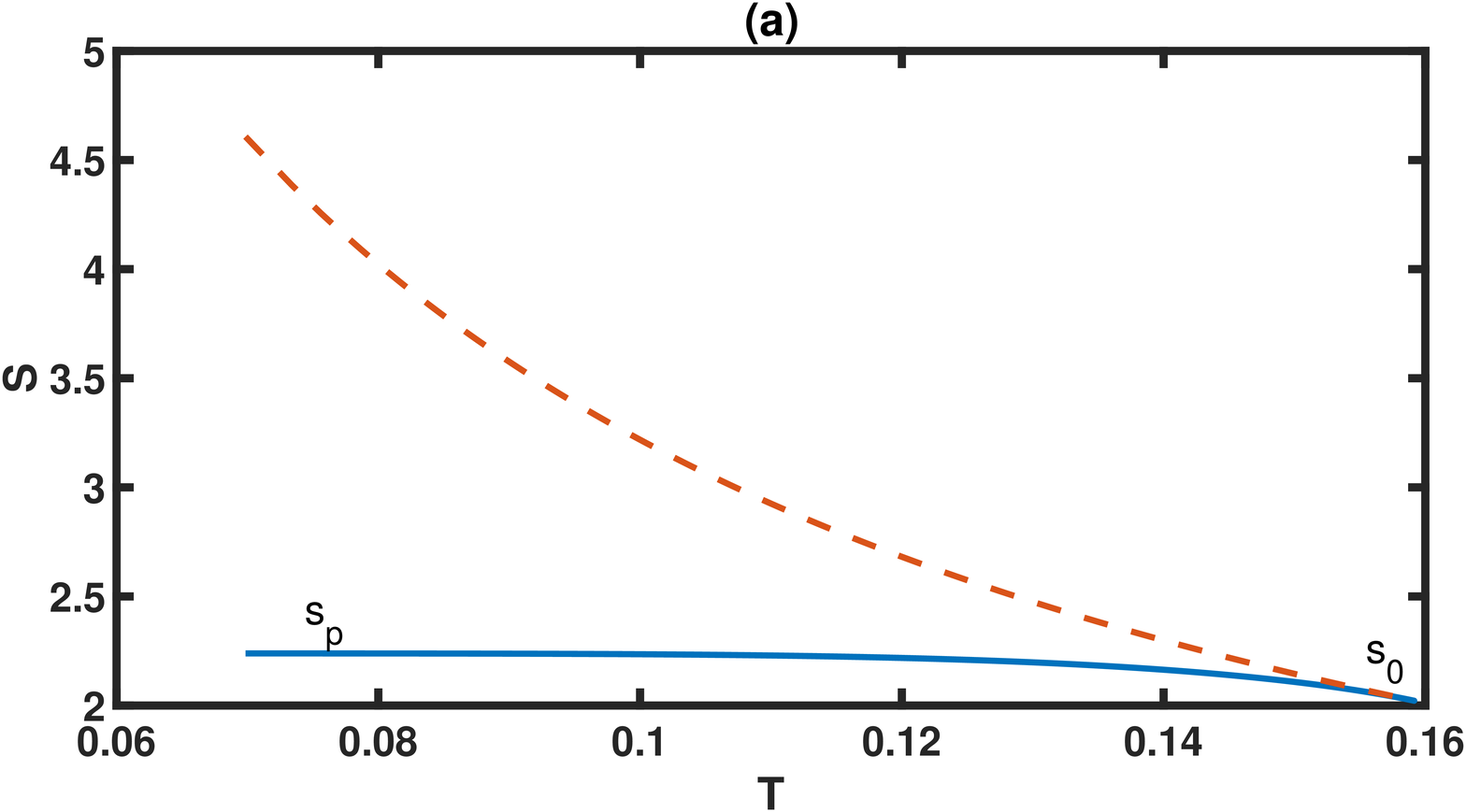}
\end{minipage}%
\begin{minipage}[c]{0.52\textwidth} 
\includegraphics[width=3.in,height=2.5in]{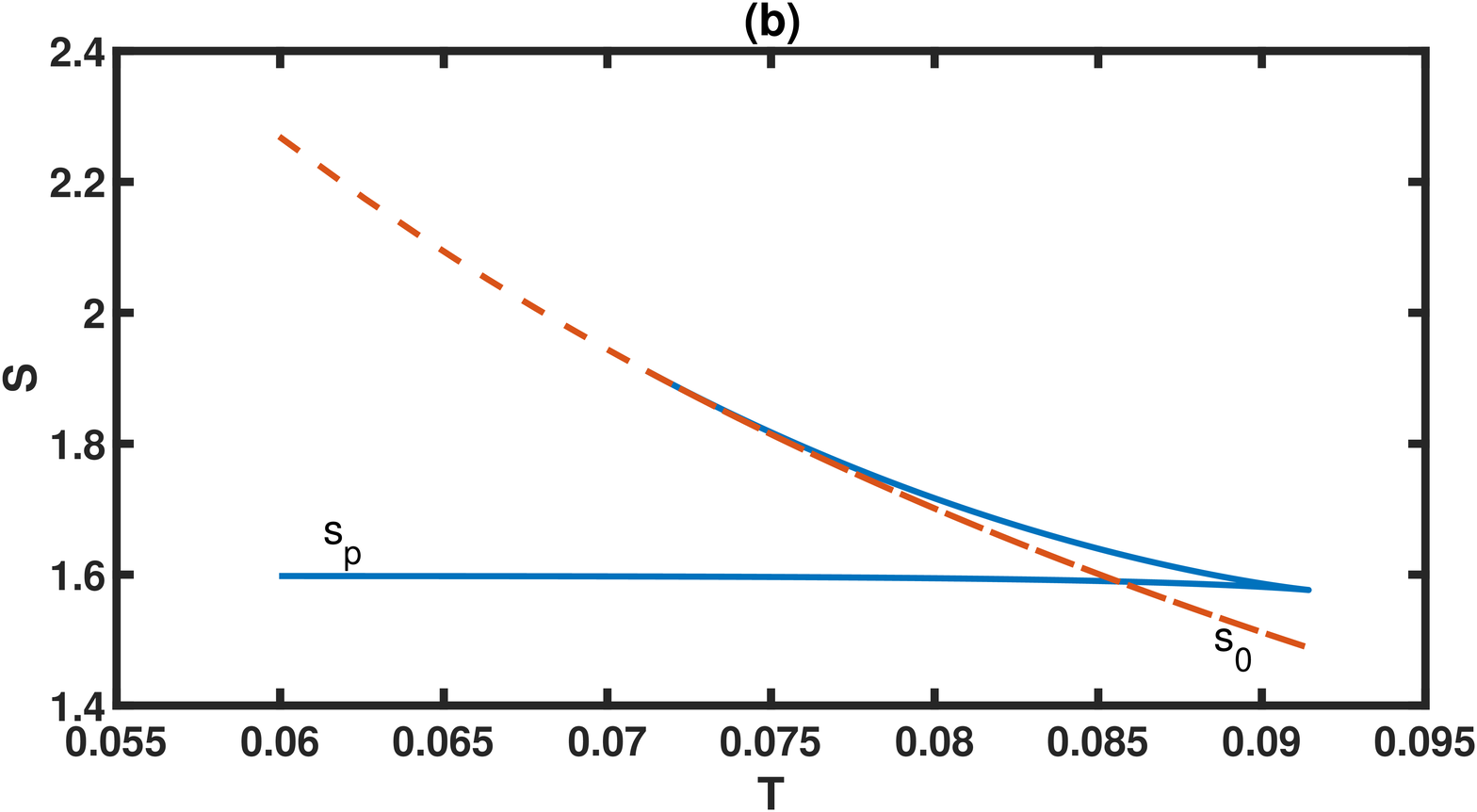}
\end{minipage}
\caption{(Colour online) The actions versus temperature for model 2, the dashed line corresponds to the thermodynamic action and the solid line to the periodon action.  (a) $\mu=1$: signature of second-order transition from quantum to thermal regimes,  (b) $\mu=3$: signature of first-order transition from quantum to thermal regimes. $a_0 = 0.5$.} 
\label{fig6}
\end{figure*}
\begin{figure*}\centering
\begin{minipage}[c]{0.52\textwidth} 
\includegraphics[width=3.in,height=2.5in]{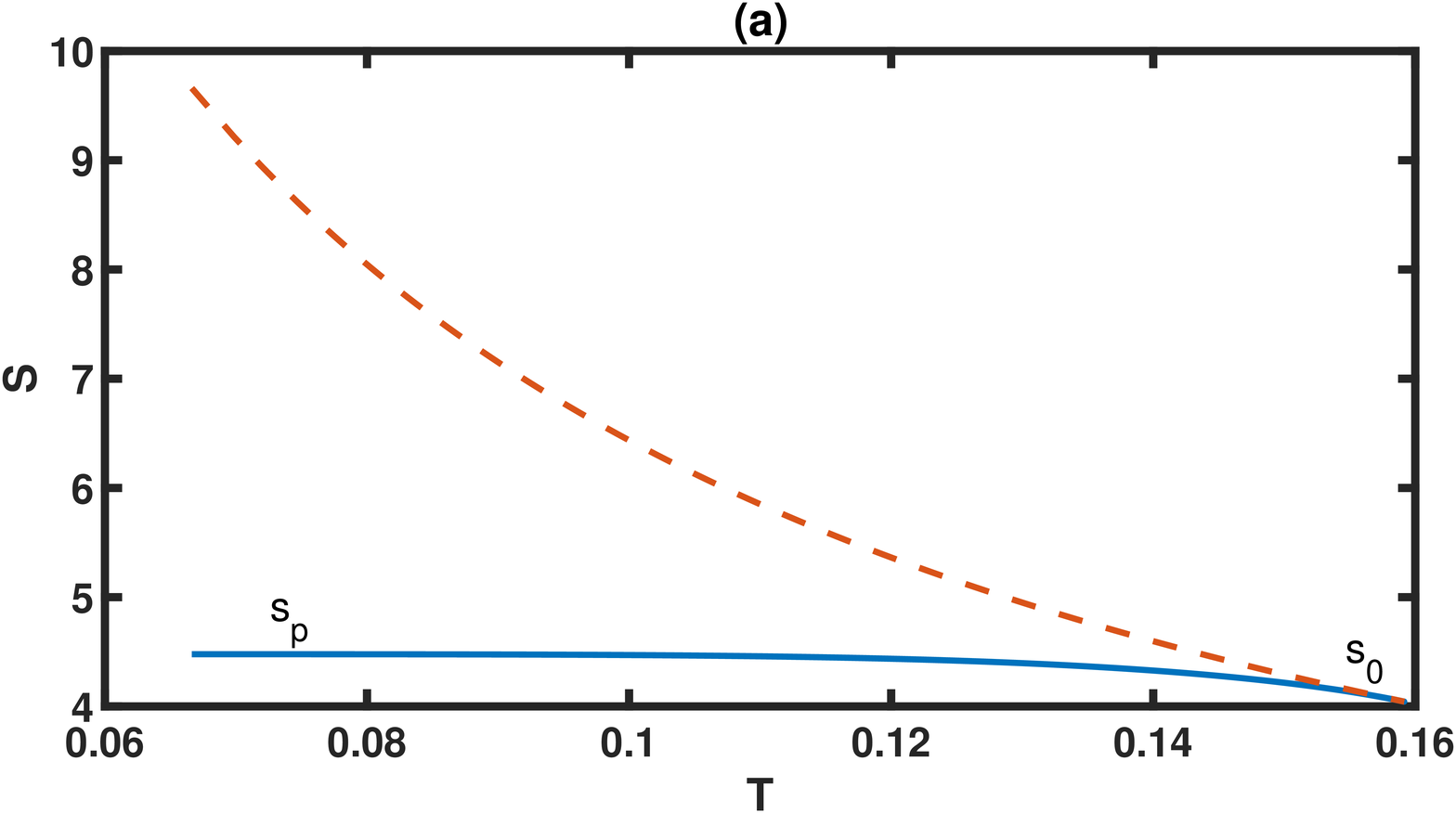}
\end{minipage}%
\begin{minipage}[c]{0.52\textwidth} 
\includegraphics[width=3.in,height=2.5in]{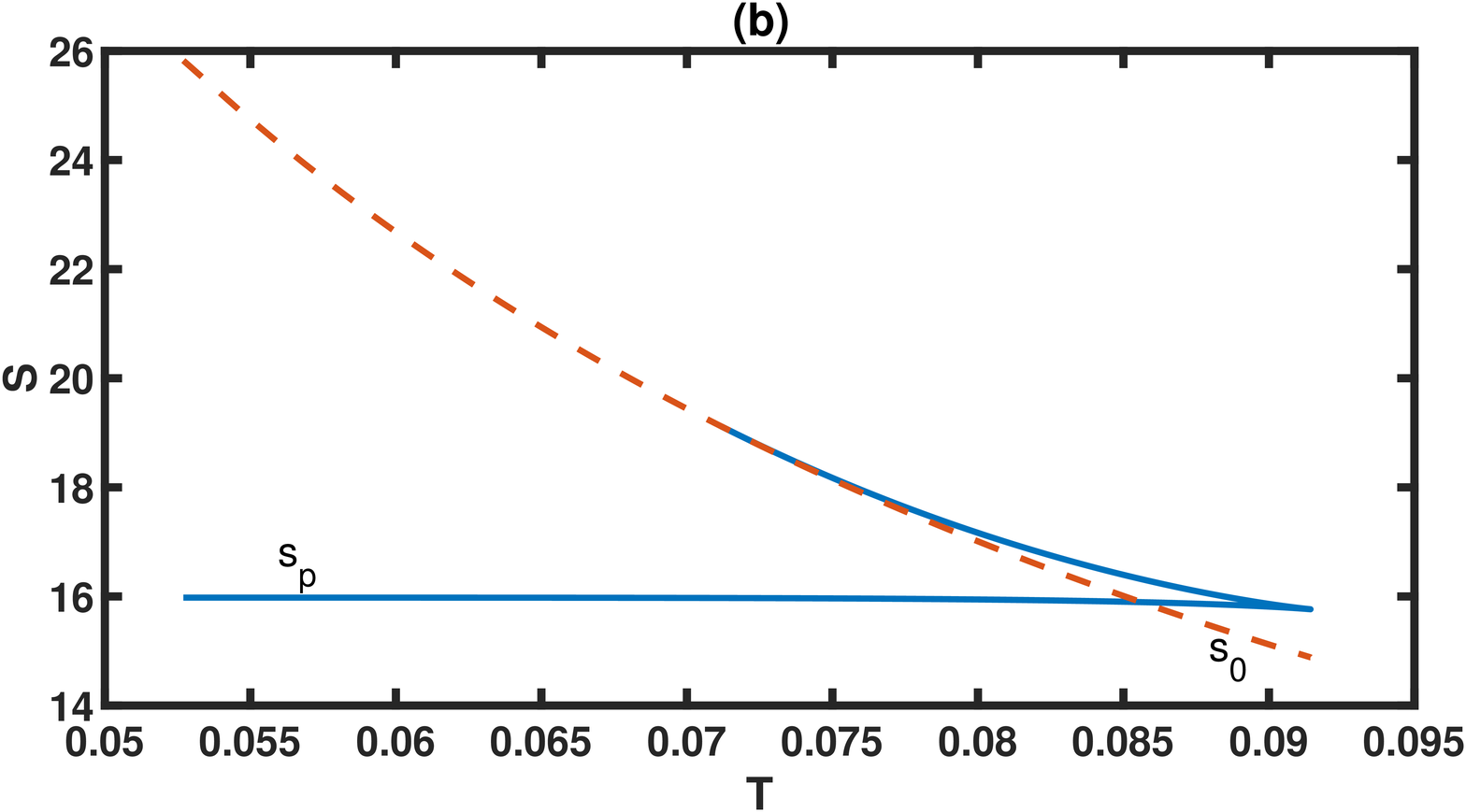}
\end{minipage}
\caption{(Colour online) The action versus temperature for model 3, the dashed line corresponds to the thermodynamic action and the solid line to the periodon action. (a) $\mu=1$: signature of second-order transition from quantum to thermal regimes, (b) $\mu=3$: signature of first-order transition from quantum to thermal regimes. $a_0 = 0.5$.} 
\label{fig7}
\end{figure*}
Characteristic features emerging from figs.\ref{fig2}, \ref{fig3} and \ref{fig4} are also reflected in the corresponding plots of the thermodynamic action and the periodon action, shown in fig. \ref{fig5}, \ref{fig6} and \ref{fig7} still for $\mu=1$ and $\mu=3$. Here too the three models display similar behaviors, indeed a smooth change from $S_{p}$ to $S_{0}$ is observed when the temperature increases for $\mu=1$ as we can see in figs. \ref{fig5}(a), \ref{fig6}(a) and \ref{fig7}(a). Clearly such change is, from the standpoint of Chudnovsky criteria, the signature of a second-order transition. Figs. \ref{fig5}(b), \ref{fig6}(b) and \ref{fig7}(b) are characterized by an abrupt change in the temperature dependence of the minimum action, so they satisfy Chudnovsky's second criterion \cite{7} for a first-order transition from quantum to thermal regimes.\par
To provide an analytical proof of a possible finite critical value of $\mu$ for a first-order transition in quantum tunneling, we follow ref. \cite{d}. We postulate that if the period $\tau_{p}(E \rightarrow a(\mu))$ of the periodon close to the barrier top can be found, then the relation $\tau_{p}(E \rightarrow a(\mu)) - \tau_{s} < 0$ (or $\omega^{2} - \omega_{s}^{2}>0$) is a sufficient condition for the system to exhibit a first-order transition. Note that $\tau_{s}$ denotes the period of small oscillations around the sphaleron, $\omega_{s}$ is the corresponding oscillation frequencies and $\omega$ is the frequency of oscillations around the energy barrier of height $a(\mu)$. For the three models this relation yields \cite{d}:
\begin{equation}
V''''(u_{sph},\mu) - \frac{5\left[ V'''(u_{sph},\mu)  \right]^{2}}{3V''(u_{sph},\mu)} < 0,
\label{e21}
\end{equation}
where $u_{sph} = 0$ corresponds to the position of the sphaleron solution. Since all the three families of DKDW potentials are symmetric we should have $V'''(0,\mu)=0$, and the above criterion (i.e. eq. (\ref{e21})) reduces to $V''''(0,\mu)<0$. Using the general expression of $V(u,\mu)$ given by formula (\ref{e1}) we finally obtain:
\begin{equation}
4\left(\frac{3}{\mu^{2}} - 2 \right)\frac{\alpha(\mu)^{4}a(\mu)}{\mu^{2}}< 0.
\label{e22}
\end{equation}
The last inequality reduces to $\mu^2> 3/2$, valid for the three families of DKDW potentials. 

\section{Exact statistical mechanics}
The aim of the transfer-integral formalism \cite{e,12,13} is to express the partition function associated with the statistical mechanics of low-dimensional systems, into a spectral problem whose solutions lead to thermodynamic quantities such as the free energy, the internal energy, the entropy, correlation functions, the correlation length and so on \cite{12}. This eigenvalue problem is represented by a linear Schr\"{o}dinger-like equation, which in some contexts do not admit exact solutions but nevertheless can be treated within the framework of approximation methods such as the WKB method \cite{12}. The $\phi^4$ is a typical non-integrable model, for which the WKB approximation has made it possible to obtain almost exact low-lying eigenmodes \cite{12}. Using the $\phi^4$ as generic model a generalized potential of the form $V(x) = [ (\eta+\xi)x^{2} - (\eta-\xi)]^{2}$, for which the Schr\"odinger equation can admit exact solutions for specific values of the characteristic parameters $\eta$ and $\varepsilon$, was proposed \cite{19}. Quite remarkably this generalized model can be connected to some existing double sinh-Gordon models \cite{33,new}, that turns out to be quasi-exactly solvable (QES) under appropriate conditions. Note that by quasi-exact solvability we mean the possibility to carry out exact partial diagonalization of the partition function. \par
Consider a one-dimensional chain of $N$ interacting atoms of identical mass $m$ (which we set to unity), placed in the field of a substrate potential $V(u,\mu)$, where $u$ stands for the displacement field of atoms to which we associate a conjugate $\pi$. The (dimensionless) Hamiltonian of such system can be written: 
\begin{equation}
H = \int dx\left[ \frac{1}{2}\pi^{2} + \frac{1}{2}(\partial_{x}u)^{2}  + V(u,\mu)\right],
\label{e23}
\end{equation}
where $V(u,\mu)$ is given by eq. (\ref{e1}). As found in the previous section, the Hamiltonian (\ref{e23}) admits single and periodic-kink soliton solutions given by (\ref{e8}) and (\ref{e9}), which we called respectively vacuum instanton and periodon. Here we focus on the thermodynamics of the DKDW models in the presence of vacuum instanton. Proceeding, our main concern is the calculation of the partition function which can be factorized as:
\begin{equation}
Z = Z_u Z_{\pi}.
\label{e24}
\end{equation}
In this factor the kinetic part is \cite{12}:
\begin{equation}
Z_{\pi}= (2\pi /\beta h^{2})^{N/2},
\label{e25}
\end{equation}
with $\beta=(k_B T)^{-1}$. The configurational part can be written \cite{12}:
\begin{equation}
Z_{u}= \sum_{n} \exp (-\beta N \epsilon_{j}),
\label{e26}
\end{equation}
where $\epsilon_{j}$ are eigenvalues of the transfer-integral operator, involving the strain (i.e. interatomic) and on-site potential (i.e. one-body) energy terms in the Hamiltonian (\ref{e23}). In the displacive regime \cite{e,12,dak} the eigenvalues $\epsilon_j$ are those of the Schr\"odinger-like equation:
\begin{eqnarray}
-\frac{1}{2\beta^{2}}\frac{\partial^{2}}{\partial u^{2}}\psi_{j} + a(\mu)\left( \frac{\sinh ^{2}(\alpha(\mu) u)}{\mu^{2}} - 1\right)^{2}\psi_{j} = \epsilon_{j}\psi_{j}. \nonumber \\
\label{e27}
\end{eqnarray}
Let us adopt the variable change $z = \alpha(\mu)\,u$ and introduce the new parameters:
\begin{equation}
\xi = (1+2\mu^{2})^{-1}, \hskip 0.25truecm
E_{j} = \frac{4\mu^{4}\xi^{2}}{a(\mu)}\epsilon_{j}.
\label{e28}
\end{equation}
With these new quantities eq. (\ref{e27}) becomes:
 \begin{eqnarray}
\frac{1}{2\beta^{2}}\frac{\partial^{2}}{\partial z^{2}}\psi_{j} &+&\frac{a(\mu) }{4\mu^{4}(\alpha(\mu)\xi)^{2}}\left[E_{j} - \left( \xi \cosh(2z) - 1\right)^{2}\right]\psi_{j} \nonumber \\ &=& 0.
\label{e29}
\end{eqnarray}
In ref. \cite{a2}, solving the eigenvalue problem for a quantum-mechanical system with a hyperbolic potential closely similar to eq. (\ref{e29}), Konwent established that provided specific constraints are imposed between characteristic parameters of the spectral equation, exact eigen solutions could exist. Following the same consideration let us introduce a constraint between the temperature $T$ and the deformability parameter $\mu$, through:
 \begin{equation}
\beta^{2} = \frac{2\mu^{4}(\alpha(\mu)\xi)^{2}}{a(\mu)}q^{2},
\label{e30}
\end{equation} 
where $q$ is assumed to be a positive integer. Substituting eq. (\ref{e30}) into eq. (\ref{e29}), we obtain \cite{a2}:
 \begin{equation}
\frac{\partial^{2}}{\partial z^{2}}\psi_{j} + q^{2} \left[E_{j} - \left( \xi \cosh(2z) - 1\right)^{2}\right]\psi_{j} = 0.
\label{e31}
\end{equation}
Eq. (\ref{e31}) describes a QES system for which eigen solutions can be obtained for $q = 1,2,3,4$, etc., each  value of $q$ defining a particular set of temperatures. Instructively when $q=1$, eq. (\ref{e31}) reduces to the spectral problem considered by Razavy \cite{20} and for which he proposed exact solutions.\\
For arbitrary integer values of $q$, solutions of eq. (\ref{e31}) vanish asymptotically as $z\rightarrow \pm \infty$. Therefore we can represent these solutions as:
\begin{equation}
\psi(z) = r(z)\exp\left[ -z_{0}\cosh(2z)\right],
\label{e32}
\end{equation}
in this representation the function $r(z)$ should be a polynomial and should be a linear combination of either $\cosh m_{p}z$ or $\sinh  m_{p}z$, where $z_{0}$ and $ m_{p}$ ( $p=0,1,2,...$) are constants to be determined. Expressions for the (unnormalized) ground states and the associated eigen energies, for the first four values of $q$, are obtained as: \par
$q=1$:
 \begin{eqnarray}\label{e33}
 \psi_{0}(u) = \exp\left[ -\frac{\cosh(2\alpha(\mu)u)}{2(1+2\mu^{2})}\right], \label{e33a} \\ 
 \epsilon_{0}(\mu) = \frac{a_{0}}{4\mu^{4}}\left[ 1 + (1+2\mu^{2})^{2}\right].
 \label{e33b}
 \end{eqnarray}
\par $q=2$:
  \begin{eqnarray}\label{e34}
  \psi_{0}(u) = \cosh(\alpha(\mu)u)\exp\left[ -\frac{\cosh(2\alpha(\mu)u)}{1+2\mu^{2}}\right], \label{e34a} \\
  \epsilon_{0}(\mu) = \frac{a_{0}}{16\mu^{4}}\left[ 3(1+2\mu^{2})^{2}-8\mu^{2}\right].
 \label{e34b}
 \end{eqnarray}
\par  $q=3$:
 \begin{eqnarray}
\psi_{0}(u) &=& \frac{6}{1+2\mu^{2}}\exp\left[ -\frac{3\cosh(2\alpha(\mu)u)}{2(1+2\mu^{2})}\right] \nonumber \\ 
  &+& \left( 1 + \sqrt{1+\frac{36}{(1+2\mu^{2})^{2}}}\right) \cosh(2\alpha(\mu)u)\nonumber \\
 &\times& \exp\left[ -\frac{3\cosh(2\alpha(\mu)u)}{2(1+2\mu^{2})}\right], \label{e35} \\
  \epsilon_{0}(\mu) &=& \frac{a_{0}}{36\mu^{4}}\left[9 -2(1+2\mu^{2})\sqrt{(1+2\mu^{2})^{2} +36}\right] \nonumber \\ &+& \frac{7a_{0}}{36\mu^{4}}(1+2\mu^{2})^{2}.
 \label{e36}
 \end{eqnarray}
 \par $q=4$:
  \begin{eqnarray}
  \psi_{0}(u) &=& \frac{12\cosh(\alpha(\mu)u)}{1+2\mu^2}\exp\left[-\frac{2\cosh(2\alpha(\mu)u)}{1+2\mu^{2}}\right] \nonumber \\ &+& 2\left(2\mu^{2}-1 +\sqrt{12-8\mu^{2}+(1+2\mu^{2})^{2}}\right) \nonumber \\ &\times& \frac{\cosh(3\alpha(\mu)u)}{1+2\mu^{2}} \exp\left[ -\frac{2\cosh(2\alpha(\mu)u)}{1+2\mu^{2}}\right],
 \label{e37} \\
  \epsilon_{0}(\mu) &=& \frac{a_0}{16\mu^4}(2 -4\mu^2) \nonumber \\ &-& \frac{a_0}{16\mu^4}(1+2\mu^2)\sqrt{(1+2\mu^2)^2 -8\mu^2 +12} \nonumber \\ &+& \frac{11a_0}{64\mu^4}(1+2\mu^2)^2.
 \label{e38}
 \end{eqnarray}
Solutions corresponding to eigenstates above the ground state, are given in the Appendix for the same four selected values of $q$.\par
\begin{figure*}\centering
\begin{minipage}[c]{0.33\textwidth} 
\includegraphics[width=2.15in,height=2.in]{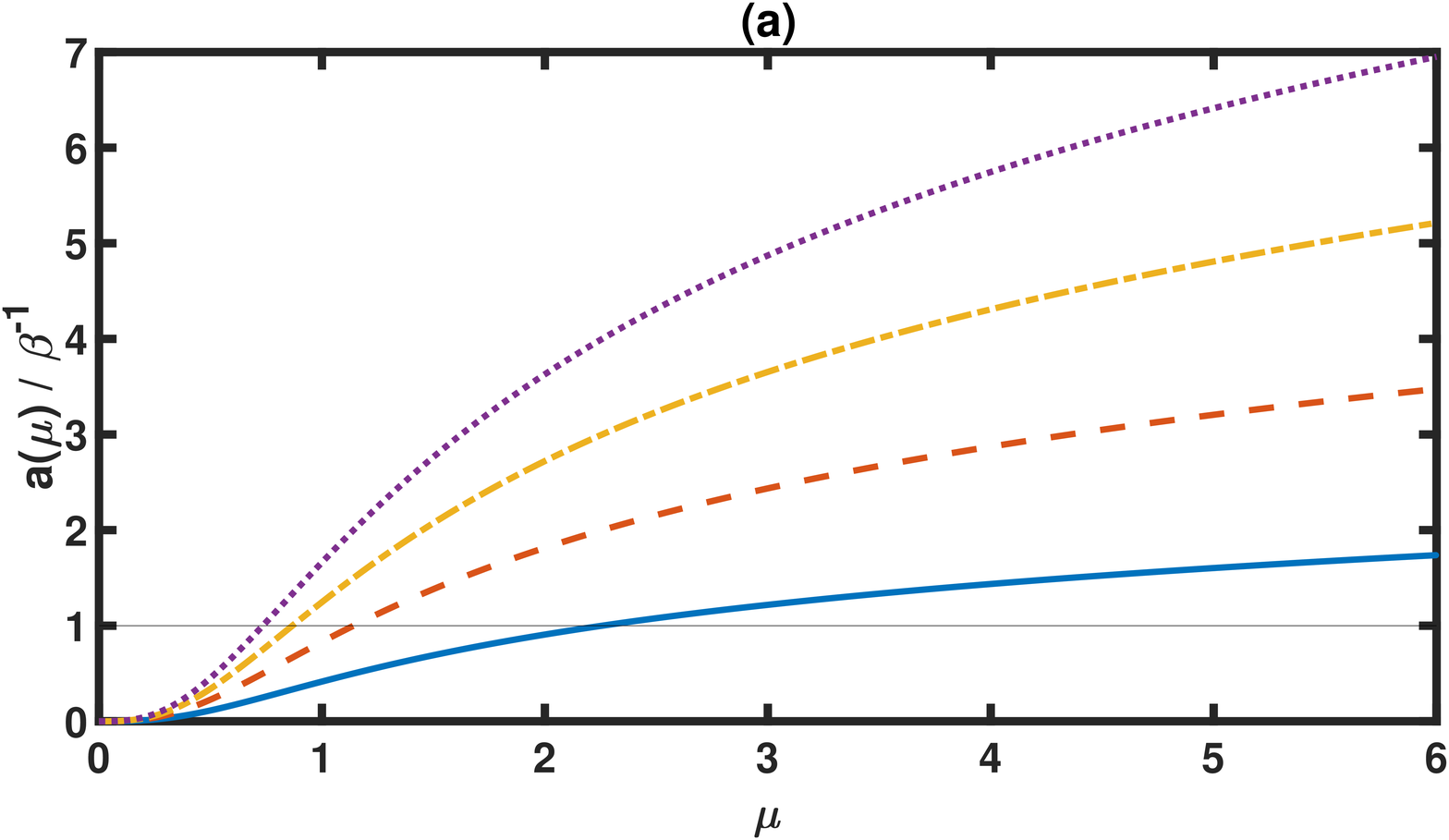}
\end{minipage}%
\begin{minipage}[c]{0.33\textwidth} 
\includegraphics[width=2.15in,height=2.in]{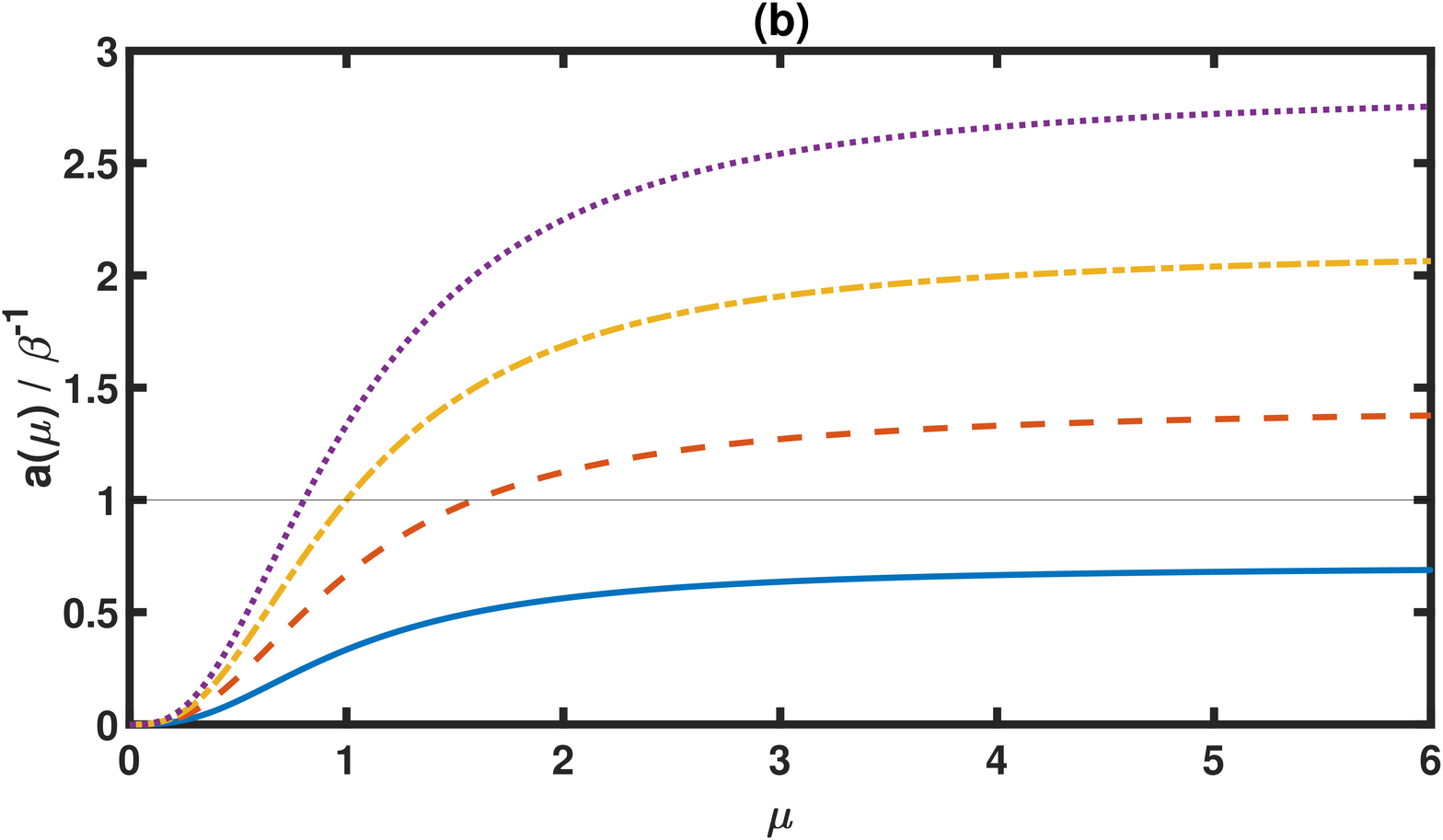}
\end{minipage}%
\begin{minipage}[c]{0.33\textwidth} 
\includegraphics[width=2.15in,height=2.in]{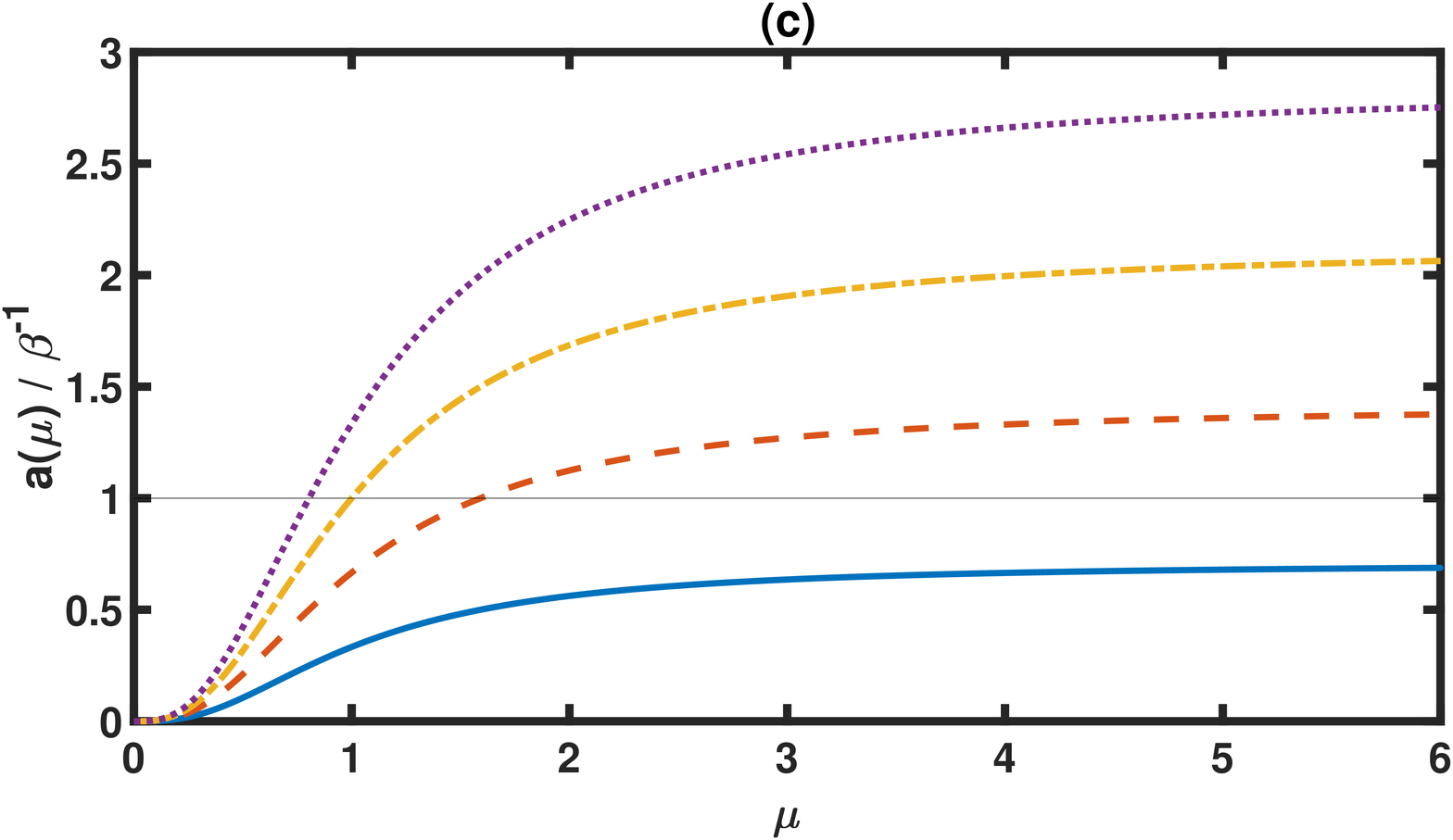}
\end{minipage}
\caption{(Colour online) Plot of the ratio of the temperature and barrier energy versus the shape deformability parameter taking  $q=1$ (Solid line), $q=2$ (Dashed line), $q=3$ (Dash-dotted line), $q=4$ (Dotted line) in (a) model 1, (b) model 2, (c) model 3. Here $a_{0}=1$ and the horizontal line mark the region where the temperatures coincide the symmetry breaking temperature.}
\label{fig8}
\end{figure*}

\begin{figure*}\centering
\begin{minipage}[c]{0.33\textwidth}  
\includegraphics[width=2.15in,height=2.in]{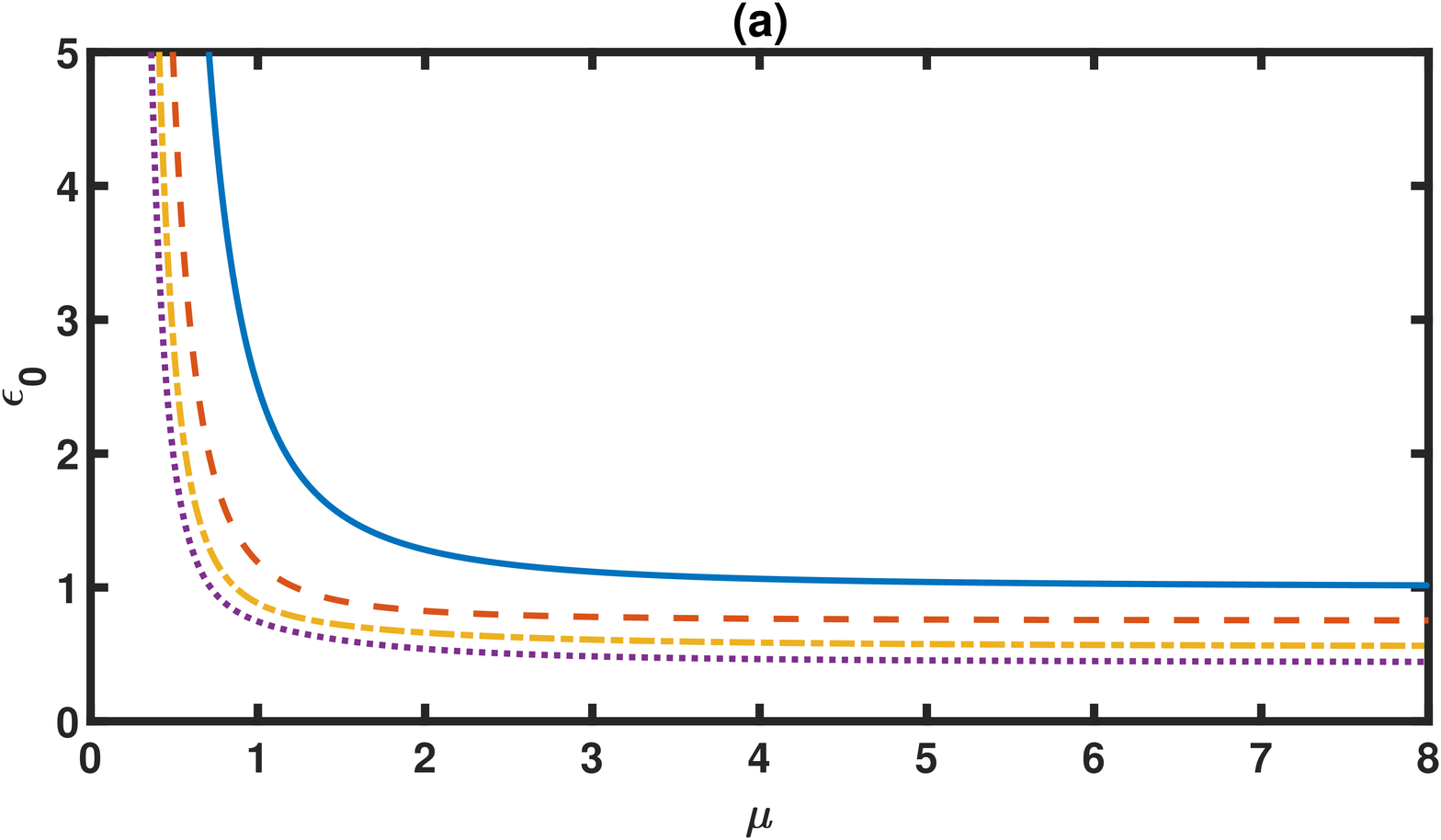}
\end{minipage}%
\begin{minipage}[c]{0.33\textwidth} 
\includegraphics[width=2.15in,height=2.in]{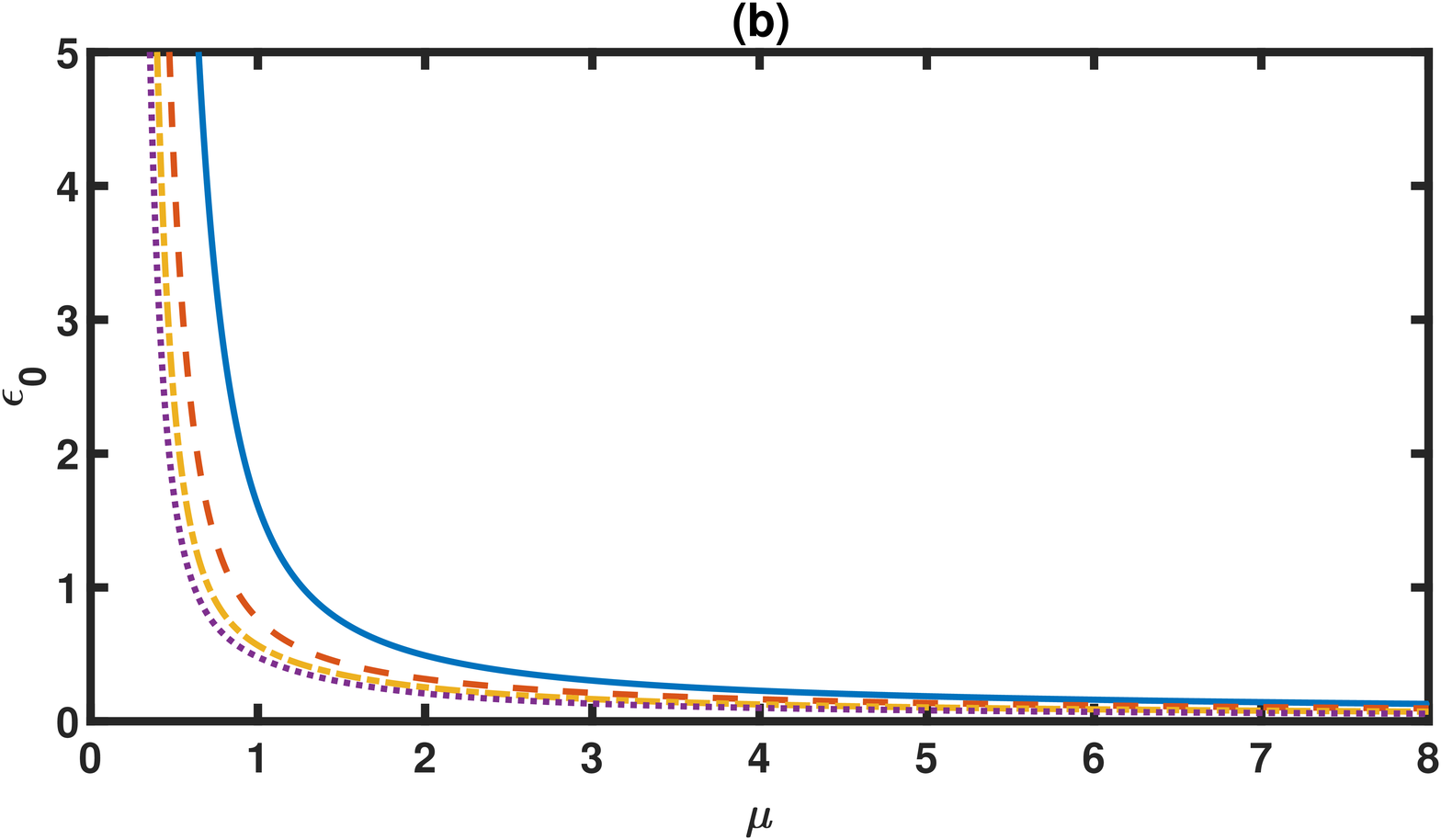}
\end{minipage}%
\begin{minipage}[c]{0.33\textwidth} 
\includegraphics[width=2.15in,height=2.in]{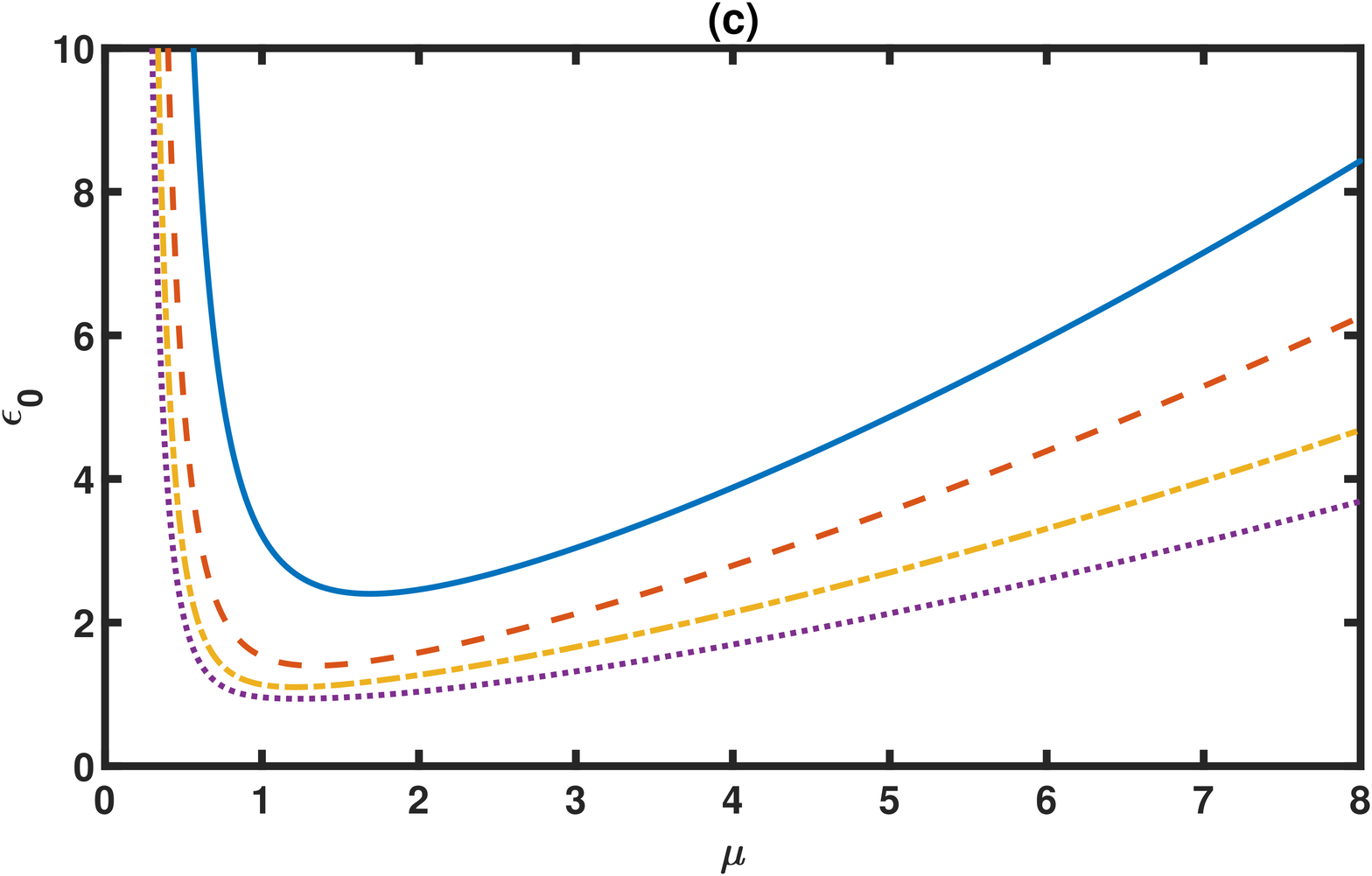}
\end{minipage}
\caption{(Colour online) Ground-state energy $\epsilon_{0}$ versus the shape deformability parameter $\mu$, for four different values of $q$ namely: $q=1$ (Solid line), $q=2$ (Dashed line), $q=3$ (Dash-dotted line), $q=4$ (Dotted line). Graph (a) is for model 1, graph (b) is for model 2, graph (c) is for model 3. $a_0=1$.}
\label{fig9}
\end{figure*}

\begin{figure}\centering
\includegraphics[width=3.5in,height=2.in]{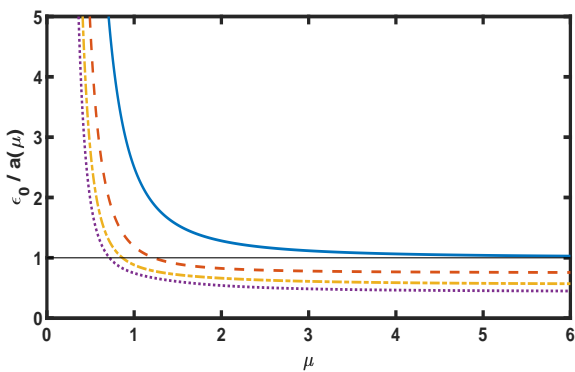}\\
\caption{(Colour online) Plot of the ratio of the ground-state energy $\epsilon_0$ to the barrier height $a(\mu)$, versus the shape deformability parameter $\mu$ for  $q=1$ (Solid line), $q=2$ (Dashed line), $q=3$ (Dash-dotted line), $q=4$ (Dotted line). Here $a_{0}=1$, and the horizontal line marks the region where the ground-state energy is located exactly on top of the barrier.} 
\label{fig10}
\end{figure}

 
Eq. (\ref{e30}) actually represents the condition for quasi-exact solvability of the system, and can be exploited to analyze the ratio $a(\mu)/ \beta^{-1}$ of the energy barrier to the thermodynamic energy $\beta^{-1}$. In fig. \ref{fig8} we plotted the ratio $a(\mu)/ \beta^{-1}$ as a function of $\mu$, for four different values of $q$. One sees that for the three families of DKDW models, the solvability requires $ \beta^{-1} \rightarrow \infty$ in the limit $\mu \rightarrow 0$. This is in agreement with known results of the transfer-integral formalism for the $\phi^{4}$ model, i.e. the spectral problem for this model does not admits exact solutions at finite temperatures \cite{12}. For small values of $\mu$ the solvability condition suggests that $k_B T>> a(\mu)$ for the three models. As $\mu$ increases, temperatures obtained from the lowest to the largest values of $q$ steadily decrease. But while going far below the energy barrier in the first model as it is noticeable in fig.\ref{fig8}(a), in fig.\ref{fig8}(b) and fig.\ref{fig8}(c) we observe that the temperatures in model 2 and model 3 instead tend to a particular limit. We found analytically that this limit is coincidentally the same for both models and $a(\mu)/ \beta^{-1}\rightarrow q/\sqrt{2}$ as $\mu \rightarrow \infty$. Therefore the solvability condition will hold at four temperatures below the symmetry breaking temperature in model 1, whereas only three temperatures below the energy barrier will meet the condition in both model 2 and model 3 given that the temperature at $q=1$, in these two last models, is always greater than the energy barrier despite its decrease with an increase of $\mu$.\\
According to the transfer-integral formalism, the free energy is intimately related to the ground-state energy $\epsilon_0$ \cite{12,dak}. The influence of the shape deformability parameter $\mu$ on the ground-state energies is represented in fig. \ref{fig9}. The three models are expected to display infinitely high ground-state energies in the limit $\mu \rightarrow 0$, in agreement with the non-integrability of the $\phi^{4}$ model within the framework of the transfer-integral formalism. As $\mu$ is increased, the ground-state energies in model 1 and model 2 drastically decrease to a critical limit as seen in fig. \ref{fig9}(a) and fig. \ref{fig9}(b), respectively. As for model 3, fig.  \ref{fig9}(c) shows that the ground-state energies corresponding to different $q$ decrease to a minimum, then steadily rise back to infinity. \\
Let us also examine the influence of $\mu$ on the competition between the ground-state energy and the barrier height $a(\mu)$, taking into consideration the variation of $q$ and hence of temperature. In fig. \ref{fig10} we plot $ \epsilon_{0}/ a(\mu)$ as a function of $\mu$, for four different values of $q$. 
The motivation for studying this ratio resides in the important fact that it does actually not depends on the choice of $a(\mu)$ and $\alpha(\mu)$, so the observations done on the relative position of the energy levels will be valid for the three models. In fig. \ref{fig10} we observe that the ground-state energy is infinitely higher than the barrier height for $\mu \rightarrow 0$, and drops with an increase of $\mu$ irrespective of the temperature. Moreover,  increasing $q$ for a fixed value of $\mu$ will yield a lower energy level. This means that the choice of $q$, combined with an increase of $\mu$, has a great impact on the position of the ground-state energy level with respect to the energy barrier. For instance, by choosing $q=1$, the energy level critically drops but will always be located above the energy barrier. A similar drop is observed when choosing higher values of $q$. However, when $\mu$ grows higher than a critical value namely $\mu_{s}= \sqrt{3/2}$, $\mu_s=\sqrt{3/4}$ and $\mu_{s}\approx 0.717$ for $q= 2$, $q=3$ and $q=4$ respectively, the energy level falls below the energy barrier and tends to a finite value and the ratio $(\epsilon_{0}/a(\mu))(\mu\rightarrow \infty) =(3/4)$ for $q = 2$, $(\epsilon_{0}/a(\mu))(\mu\rightarrow \infty)= (5/9)$ for $q= 3$   and $(\epsilon_{0}/a(\mu))(\mu\rightarrow \infty)= (7/16)$ for $q=4$.\\
\begin{figure*}\centering
\begin{minipage}[c]{0.33\textwidth}  
\includegraphics[width=2.15in,height=2.in]{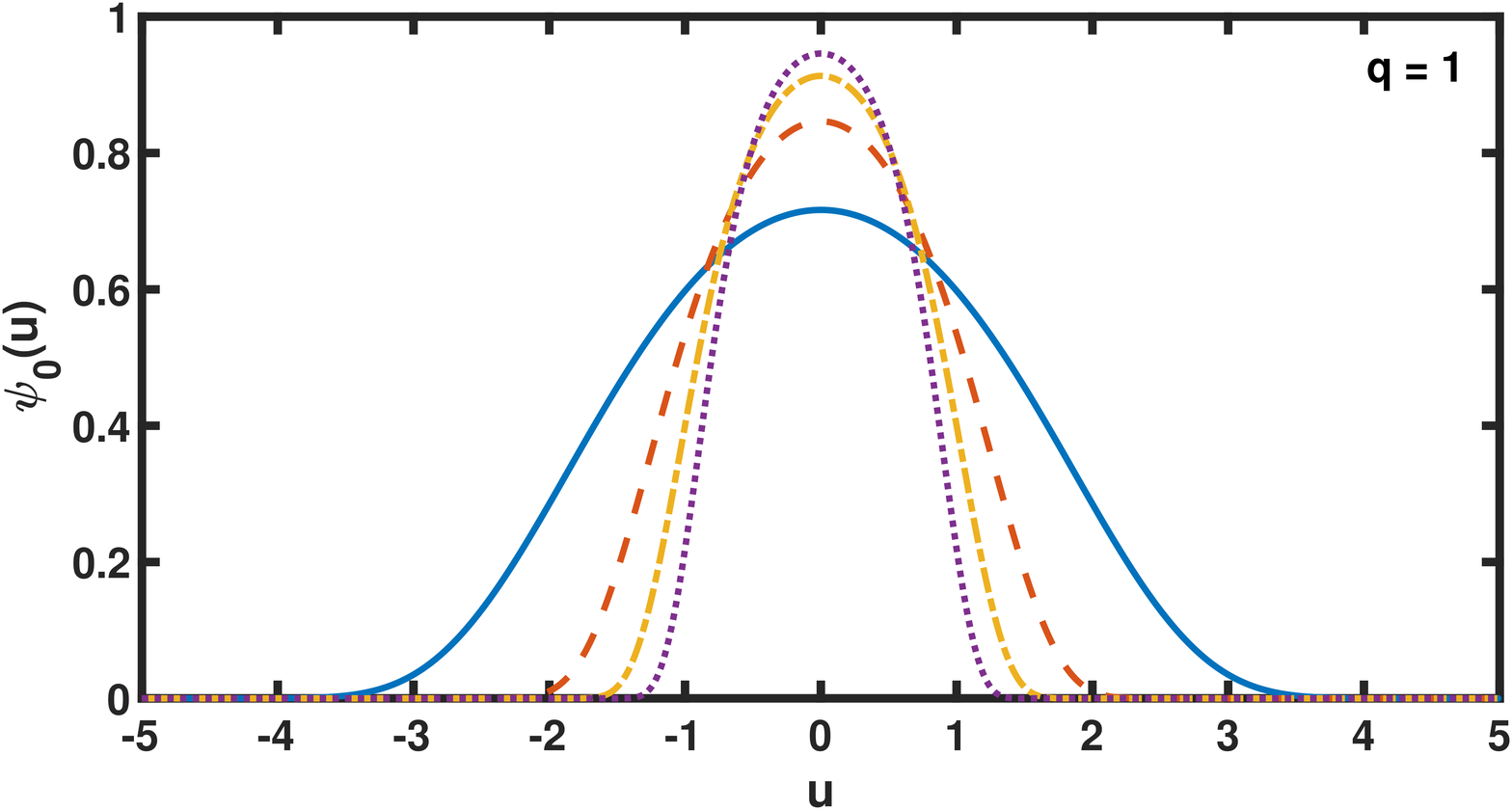}
\end{minipage}%
\begin{minipage}[c]{0.33\textwidth} 
\includegraphics[width=2.15in,height=2.in]{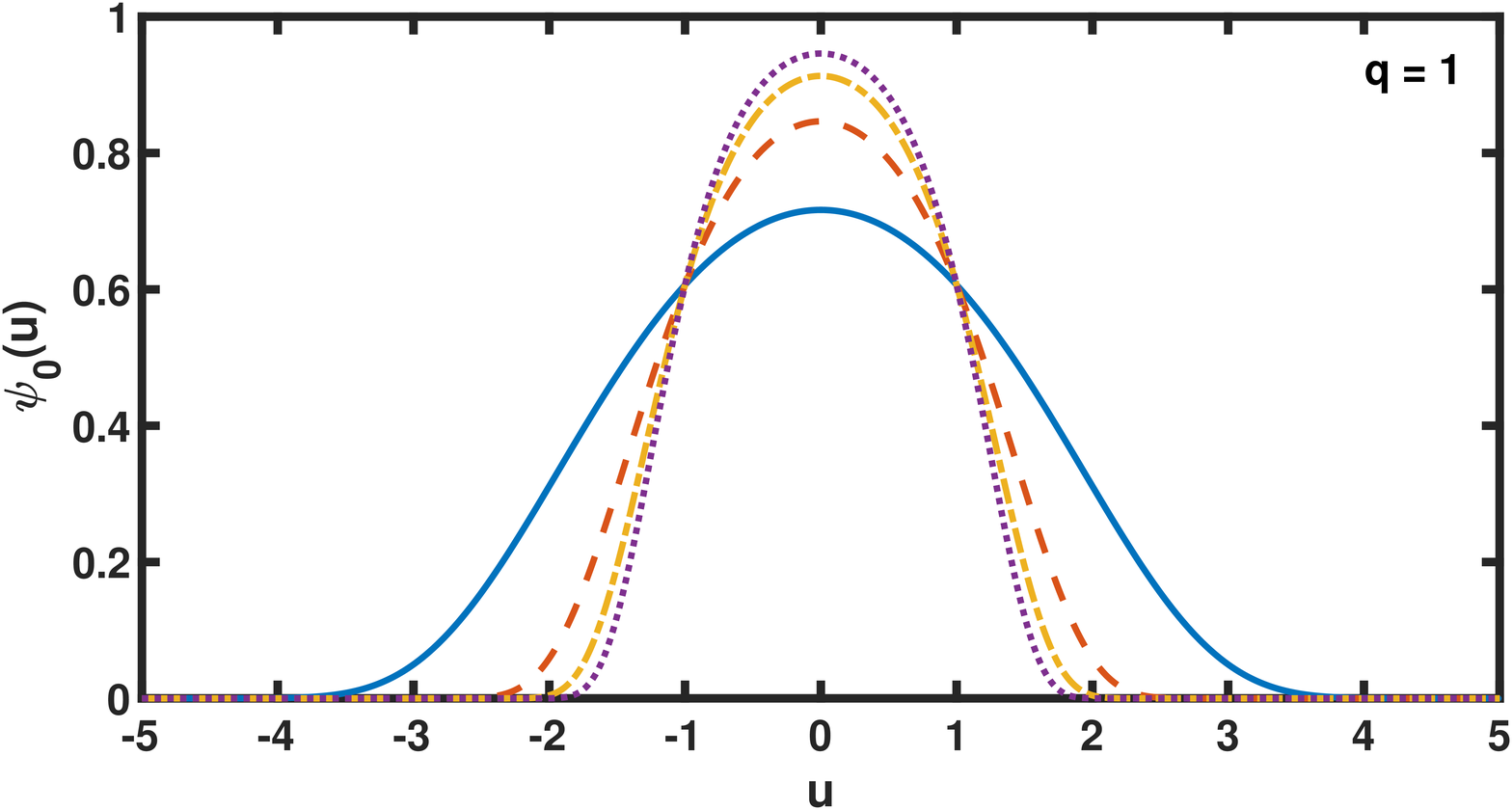}
\end{minipage}%
\begin{minipage}[c]{0.33\textwidth} 
\includegraphics[width=2.15in,height=2.in]{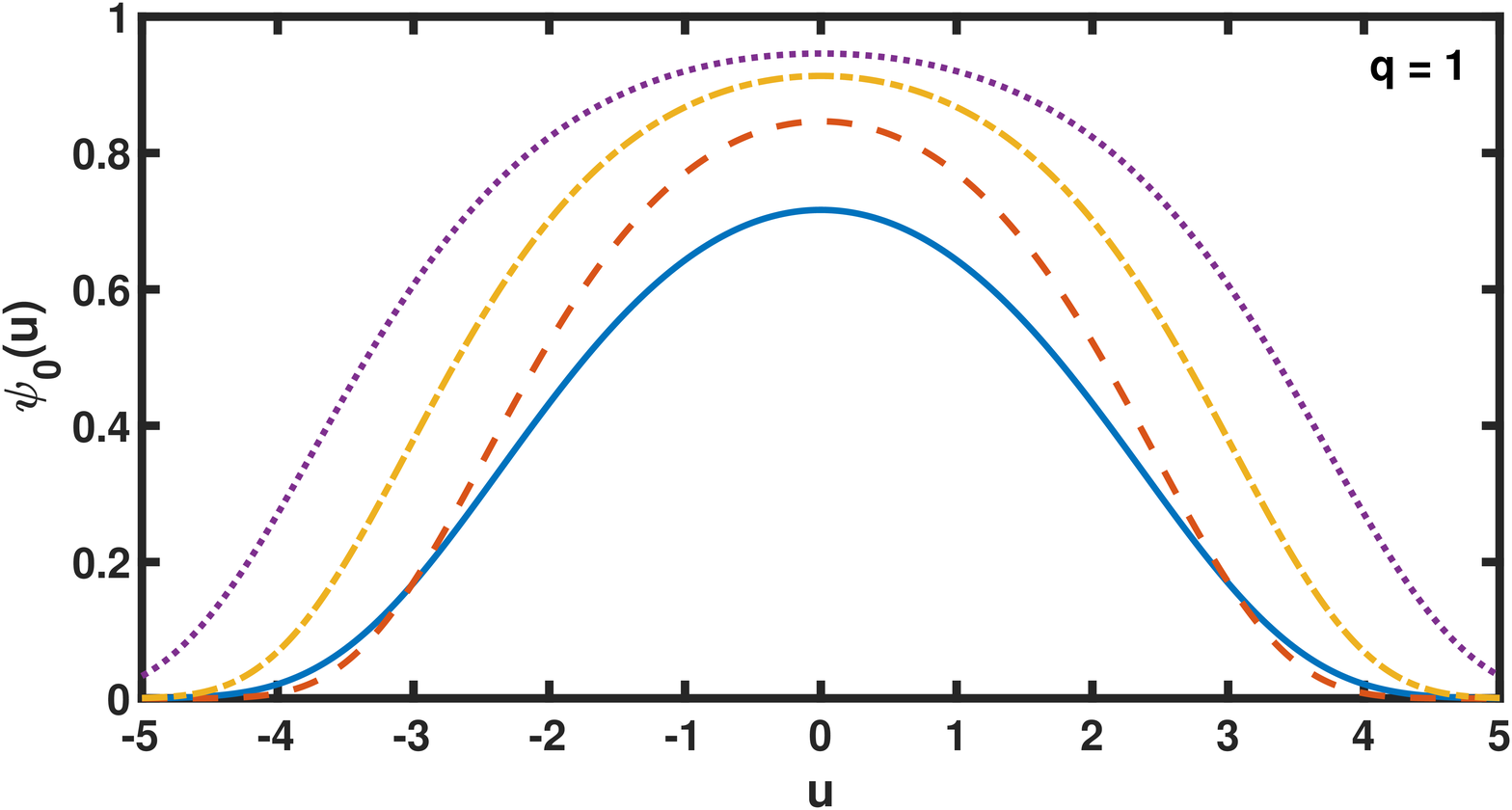}
\end{minipage}\\
\begin{minipage}[c]{0.33\textwidth} 
\includegraphics[width=2.15in,height=2.in]{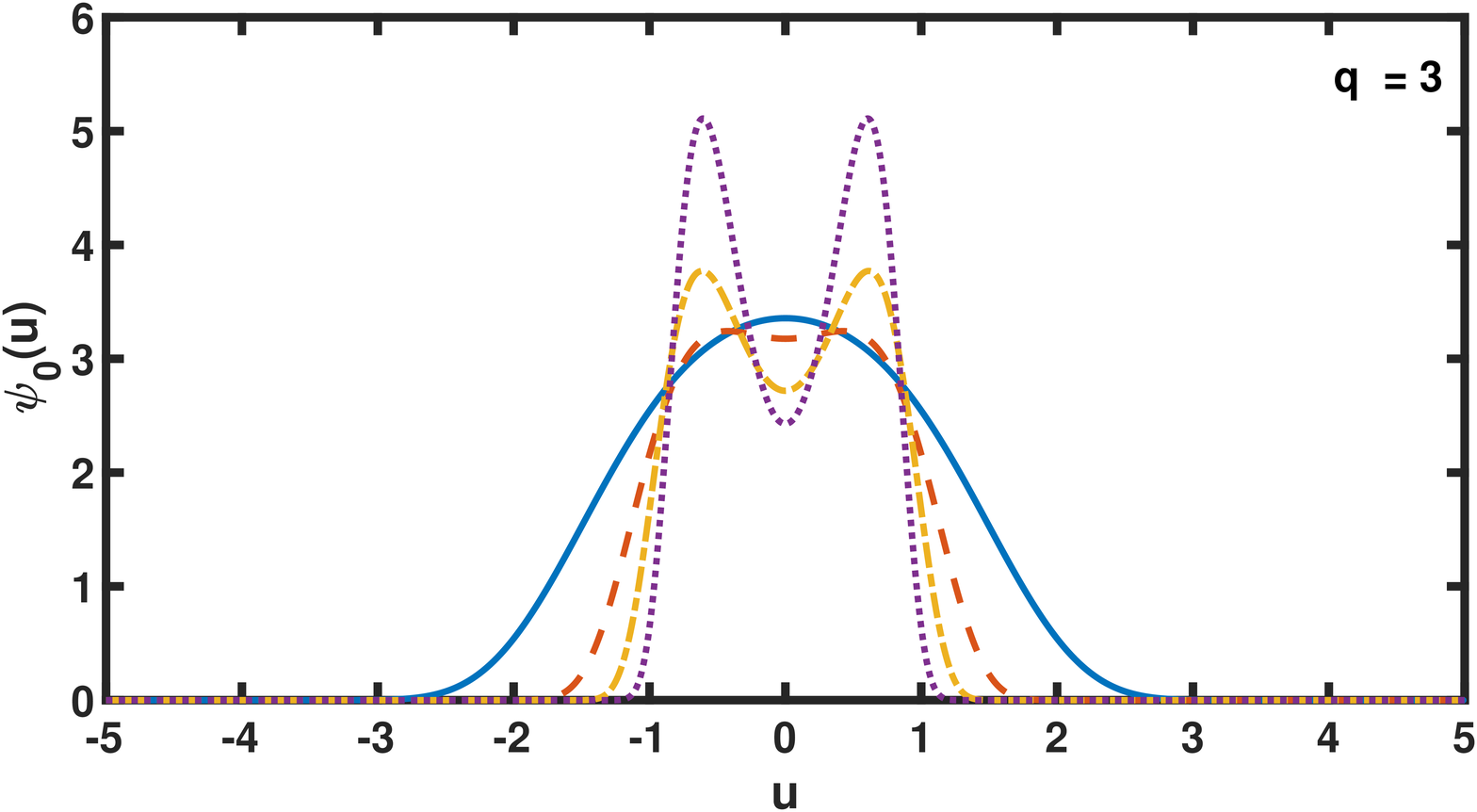}
\end{minipage}%
\begin{minipage}[c]{0.33\textwidth} 
\includegraphics[width=2.15in,height=2.in]{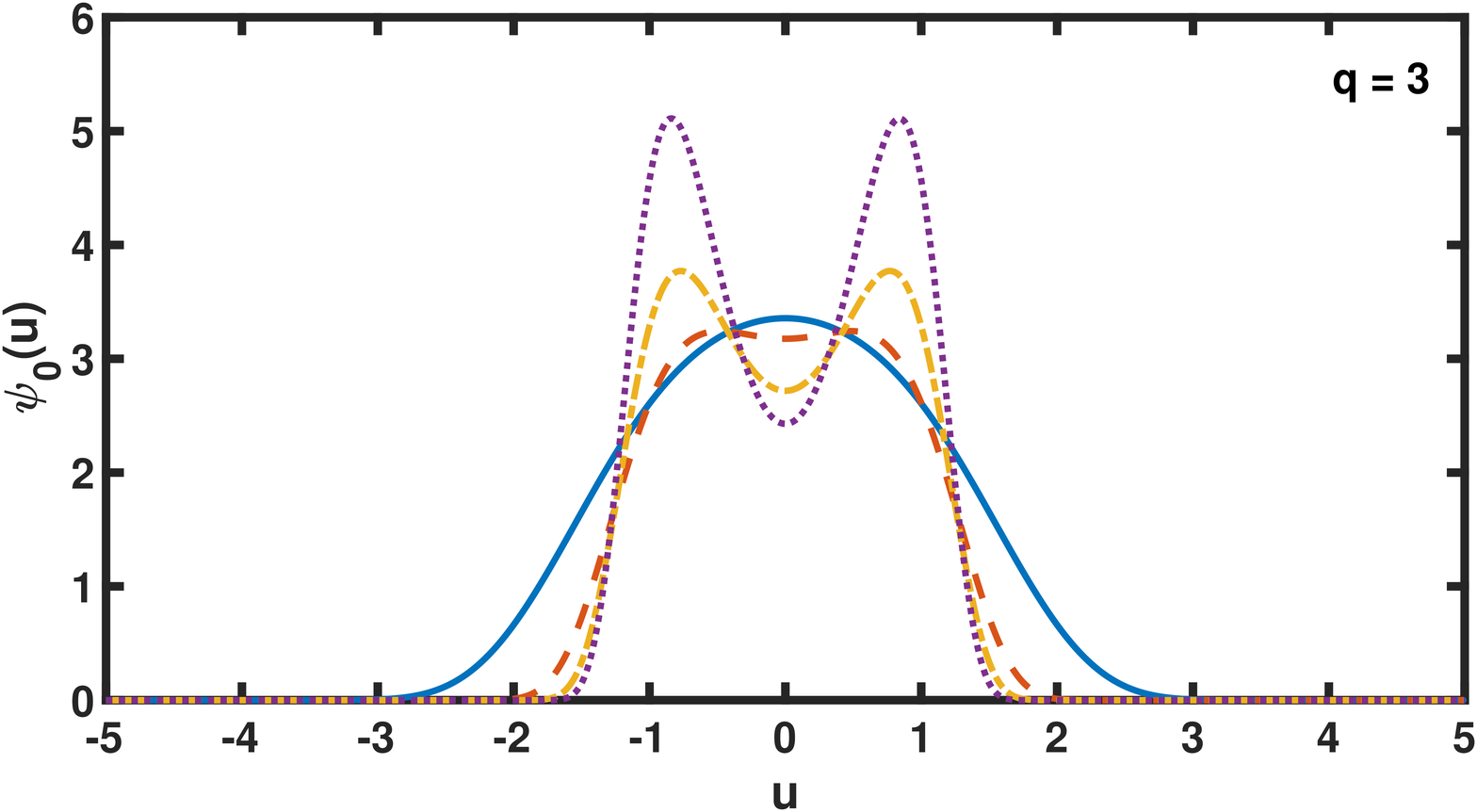}
\end{minipage}%
\begin{minipage}[c]{0.33\textwidth} 
\includegraphics[width=2.15in,height=2.in]{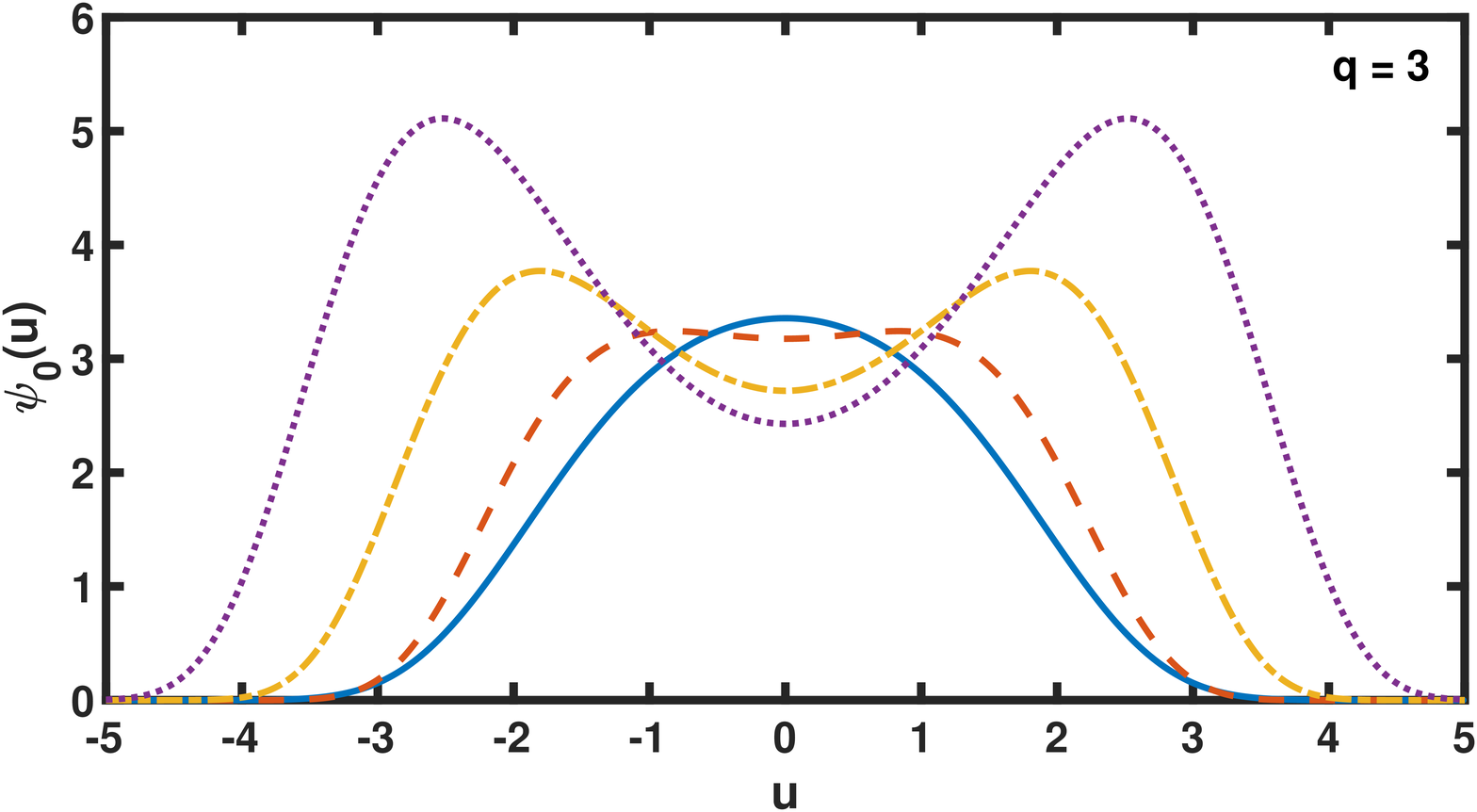}
\end{minipage}
\caption{(Colour online) Ground-state wavefunctions in position space at three temperatures, for $\mu =0.5$, for four different values of $q$(corresponding to four different temperatures ) namely: $\mu =0.5$ (Solid line),$\mu =1$ (Dashed line), $\mu =1.5$ (Dash-dotted line), $\mu =2$ (Dotted line). From left to right: model with variable position of minima, model with variable barrier height, model with variable barrier height and position of minima}
\label{fig11}
\end{figure*}
Knowing the exact ground-state energies and the corresponding eigenfunctions, it is motivating to illustrate their use in the evaluation of some thermodynamic functions. In this goal we consider a relevant quantity that plays the key role in the calculation of correlation functions \cite{12,dak}, namely the probability density. Since the wavefunctions corresponding to eigenstates of the transfer-integral operator are all real, the probability density associated with the classical field $u$ will be nothing else but the square of the normalized ground-state wavefunction. Sketched in fig. \ref{fig11} are ground-state wavefunctions for the three families of DKDW models, computed for two different temperatures and considering several values of the shape deformability parameter $\mu$. To understand the physics in the curves of fig. \ref{fig11} as concerns their variations with the deformability parameter $\mu$, it is useful to recall that in fig. \ref{fig10} we have seen that for small values of $\mu$ the ground-state energies were in the high-temperature region, as they were found to be greater than the height of the potential barrier for the three models. Consequently in this regime a pseudoparticle will possess sufficient energy to escape from one well to another over the barrier. Thus the whole configurational space becomes probable and the full state space is covered with power-law tails, with a maximum probability density at the barrier peak. As $\mu$ increases the probability density increases in amplitude for the three models, but while its width decreases for the first and second model the width of the probability density is  instead increasing for the third model. An increase of $\mu$ in the case of model 1 and model 2 causes a decrease in the width and height of the potential barrier. In the two models the shape deformability will also strengthen the steepness of the reflective walls, and thus restricts the attainable space to the region covered by just the valleys and the energy barrier. In model 3, the potential barrier increases with its width gradually increasing as the degenerate minima move far from each other. Increasing the deformability parameter $\mu$ consequently broadens the attainable space covered by the energy barrier. The ground-state energies being high in the three models at large $\mu$, the energy barrier thus remains the most probable. \par
To end this section it is relevant to stress that when analyzing properties of the probability density in the present specific context, it is relevant to always keep in mind the importance of the choice of the value of $q$. Taking $q=1$ for instance, the ground-state energy is always above the energy barrier irrespective of $\mu$. Indeed even if an increase of $\mu$ lowers the temperature the probability density will always be single picked, with the peak position coinciding with the position of the potential barrier (top graphs in fig. \ref{fig11}). For values of $q$ greater than one, an increase of $\mu$ above $\mu_{s}$ lowers the ground-state level below the energy barrier. The transitions from one well to another thus become less probable as the temperature decreases below the transition temperature. In this latter case the pseudoparticle stays much longer confined in the potential wells, such that the valleys become more probable. This trend is illustrated in fig.\ref{fig11} by the probability density exhibiting a double-peak shape, with each peak located in the vicinity of one of the two degenerate potential minima.

\section{Conclusion}
The decay of metastable states over an energy barrier is a physical process inherent to a large number of systems, and particularly those undergoing phase transitions \cite{bon1,bon2,bon3}. It is observed in the motion of atoms trapped in the field of force of a multi-well potential \cite{bon2}, the motion of kinks and dislocations across the Peierls-Nabarro barrier \cite{pn1,pn2}, the excitations of Cooper pairs in a Josephson loop \cite{lar}, the dynamics of macromolecules in DNA strands and other molecular chains across hydrogen bridges and so on. In a pioneer study Kramers \cite{bon1} suggested that the lifetime of a classical particle in a metastable state, separated from a stable equilibrium state by an energy barrier, would obey Arrhenius law. Later on it was established that Kramers theory is valid main at high temperatures, where thermal activations are likely to overcome high potential barriers. At low enough temperatures thermal fluctuations are not enough to favor a jump over the potential barrier, and only by quantum tunneling the particle will cross the barrier. It is known \cite{c} that there exists a critical temperature $T_c$ at which there is a transition from classical to quantum decay regimes. The transition can be of first order with a discontinuous first derivative of the Euclidean action, or of second-order with only a second derivative of the action developing a jump. This transition, also referred to as transition in quantum tunneling, has recently been investigated for some physical systems \cite{10,4a,a,5,6,7,8,9,11}. For systems described by the $\phi^4$ model, the transition in quantum tunneling is strictly of second order \cite{10,4a,a,c}.\par
In this study we determined conditions for the occurrence of first-order transitions in quantum tunneling in bistable systems, using a class of three double-well potential models whose shapes can be tuned by varying a deformability parameter. The first-order transition supplements the second-order transition in quantum tunneling predicted with the $\phi^4$ model, which turns out to be a specific limit of the three parametrized double-well models. The first-order transition occurs at a finite critical value of the shape deformability parameter, which is the same for all three models. We have also examined conditions for exact integrability of the partition function associated with the statistical mechanics of the three models, within the framework of the transfer-integral operator formalism. In this formalism, the partition function is mapped onto a linear Schr\"odinger equation whose eigenvalues and eigenfunctions contribute to the formulation of relevant thermodynamic quantities such as the free energy, the correlation length and the correlation functions. We established that when the thermodynamic temperature and the deformability parameter are connected by a specific relation, exact solutions to this spectral problem can exist. We derived the eigenvalues and eigenfunctions of some of these exact solutions analytically, and discussed the physical implications of their dependence on the shape deformability parameter. \par
Bistable processes are among the most frequent physical phenomena observed in nature \cite{bi1}, they usually emerge in form of a transition between two states of equivalent energy across an potential barrier and are present in sturctural phase transitions, dynamical properties of anharmonic electromechanical oscillators displaying hysteresis features \cite{des}, organized cellular structures in tissues and embryonic cells \cite{q1,q2}, glycolitic oscillations in suspensions of yeast cells in unison \cite{q1,q2}, the assembly of pacemaker cells in the sino-atrial node and so on \cite{des}. Most commonly the bistability in these systems are represented by the $\phi^4$ model, a universal model which unfortunately suffers from a weakness due to its rigid double-well profile. Fofr instance this model predicts that bistable ssystems can undergo only a second-order transition in quantum tunneling, whereas experiments have established that several physical systems with double-well energy landscapes could also undergo a first-order transition in quantum tunneling. Moreover, the transfer-integral formalism for the $\phi^4$ generates an eigenvalue problem which is not exactly solvable. The present study emphasizes the need for parametrizing the standard $\phi^4$ potential, in order to account for several physical processes observed experimentally but which the $\phi^4$ has not been able to predict. We note that the influence of the double-well shape profile of the bistable potential, on the order of transitions in quantum tunneling in multistate systems, has already been discussed in some previous works \cite{c}. Our study provides a more comprehensive insight onto this problem, given that the shapes of the three families of double-well potentials can be tuned distinctively, and consequently represent a very large variety of bistable physical systems. 

\begin{acknowledgements}
The work of A. M. Dikand\'e was carried out at the Max-Planck Institute for the Physics of Complex Systems (MPIPKS), Dresden Germany, within the framework of the "Return Fellowship" program of the Alexander von Humboldt Stiftung. 
\end{acknowledgements}

\section*{Authors contribution}
All the authors contributed equally to this work.

\appendix
\section*{Appendix}
In this appendix, we give the analytical expressions of some exact solutions of the {\bf Schr\"odinger} equation (\ref{e31}), beside those written in eqs. (\ref{e33}) to (\ref{e38}).  

\par
$q=1$:
 \begin{eqnarray}
 \psi_{0}(u) &=& \exp\left[ -\frac{\cosh(2\alpha(\mu)\,u)}{2(1+2\mu^{2})}\right], \nonumber \\ 
 \epsilon_{0}(\mu) &=& \frac{a(\mu)}{4\mu^{4}}\left[ 1 + (1+2\mu^{2})^{2}\right]
 \label{e41}
 \end{eqnarray}
\par
 $q=2$:
 \begin{eqnarray}
 \psi_{0}(u) &=& \cosh(\alpha(\mu)\,u)\exp\left[ -\frac{\cosh(2\alpha(\mu)\,u)}{1+2\mu^{2}}\right], \nonumber \\ \epsilon_{0}(\mu) &=& \frac{a(\mu)}{16\mu^{4}}\left[ 3(1+2\mu^{2})^{2}-8\mu^{2}\right] 
 \label{e42}
 \end{eqnarray}
\begin{eqnarray}
 \psi_{1}(u) &=& \sinh(\alpha(\mu)\,u)\exp\left[ -\frac{\cosh(2\alpha(\mu)\,u)}{1+2\mu^{2}}\right], \nonumber \\ \epsilon_{1}(\mu) &=& \frac{a(\mu)}{16\mu^{4}}\left[ 3(1+2\mu^{2})^{2}+8(1+\mu^{2})\right].
 \end{eqnarray}
\par
$q=3$:
 \begin{eqnarray}
 \psi_{0}(\phi) &=& \frac{6}{1+2\mu^{2}}\exp\left[ -\frac{3\cosh(2\alpha(\mu)\,u)}{2(1+2\mu^{2})}\right]\nonumber \\ &+& \left( 1 + \sqrt{1+\frac{36}{(1+2\mu^{2})^{2}}}\right) \cosh(2\alpha(\mu)\,u)\nonumber \\
 &\times& \exp\left[ -\frac{3\cosh(2\alpha(\mu)\,u)}{2(1+2\mu^{2})}\right], 
 \label{e44} \\
 \epsilon_0(\mu) &=& \frac{a(\mu)}{36\mu^{4}}\left[9 -2(1+2\mu^{2})\sqrt{(1+2\mu^{2})^{2} +36}\right] \nonumber \\ &+& \frac{7a(\mu)}{36\mu^{4}}(1+2\mu^{2})^2.
 \label{e45}
 \end{eqnarray}
 \begin{eqnarray}
 \psi_{1}(u) &=& \sinh(2\alpha(\mu)\,u)\exp\left[ -\frac{3\cosh(2\alpha(\mu)\,u)}{2(1+2\mu^{2})}\right], \nonumber \\ \epsilon_{1}(\mu) &=& \frac{a(\mu)}{36\mu^{4}}\left[ 9 + 5(1+2\mu^{2})^{2}\right],
 \label{e46}
\end{eqnarray}
 
 \begin{eqnarray}
 \psi_{2}(u) &=& \frac{6}{1+2\mu^{2}}\exp\left[ -\frac{3\cosh(2\alpha(\mu)\,u)}{2(1+2\mu^{2})}\right] \nonumber \\ &-& \left(\sqrt{1+\frac{36}{(1+2\mu^{2})^{2}}}-1\right) \cosh(2\alpha(\mu)\,u)\nonumber \\ &\times& \exp\left[-\frac{3\cosh(2\alpha(\mu)\,u)}{2(1+2\mu^{2})}\right],
 \label{e47} \\
 \epsilon_{2}(\mu) &=& \frac{a(\mu)}{36\mu^{4}}\left[9 +2(1+2\mu^{2})\sqrt{(1+2\mu^{2})^{2} +36}\right]\nonumber \\ &+& \frac{7a(\mu)}{36\mu^{4}}(1+2\mu^{2})^{2}.
 \label{e48}
 \end{eqnarray}
\par 
$q=4$
 \begin{eqnarray}
 \psi_{0}(u) &=& \frac{ 12\cosh(\alpha(\mu)\,u)}{1+2\mu^{2}}\exp\left[-\frac{2\cosh(2\alpha(\mu)\,u)}{1+2\mu^{2}}\right]\nonumber \\ &+& 2\left(2\mu^{2}-1 +\sqrt{12-8\mu^{2}+(1+2\mu^{2})^{2}}\right)\nonumber \\ &\times& \frac{\cosh(3\alpha(\mu)\,u)}{1+2\mu^2} \exp\left[ -\frac{2\cosh(2\alpha(\mu)\,u)}{1+2\mu^{2}}\right],
 \label{e49} \\
 \epsilon_{0}(\mu) &=& \frac{a(\mu)}{8\mu^{4}}(1 -2\mu^{2}) + \frac{11a(\mu)}{64\mu^{4}}(1+2\mu^{2})^{2} \nonumber \\&-&
 \frac{a(\mu)}{16\mu^{4}}(1+2\mu^{2})\sqrt{(1+2\mu^{2})^{2} -8\mu^{2} +12},
 \label{e50}
 \end{eqnarray}
 \begin{eqnarray}
 \psi_{1}(u) &=& \frac{12\sinh(\alpha(\mu)\,u)}{1+2\mu^{2}}\exp\left[ -\frac{2\cosh(2\alpha(\mu)\,u)}{1+2\mu^{2}}\right]\nonumber \\ &+& 2\left(2\mu^{2}+3 +\sqrt{20+8\mu^{2}+(1+2\mu^{2})^{2}}\right)\nonumber \\ & \times& \frac{\sinh(3\alpha(\mu)\,u)}{1+2\mu^{2}}\exp\left[-\frac{2\cosh(2\alpha(\mu)\,u)}{1+2\mu^{2}}\right],
 \label{e51}\\
 \epsilon_{1}(\mu) &=& \frac{a(\mu)}{8\mu^{4}}(3 +2\mu^{2}) + \frac{11a(\mu)}{64\mu^4}(1+2\mu^2)^2\nonumber \\ &-& \frac{a(\mu)}{16\mu^4}(1+2\mu^{2})\sqrt{(1+2\mu^{2})^{2} +8\mu^{2} +20},
 \label{e52}
 \end{eqnarray}
 and
 \begin{eqnarray*}
 \psi_{2}(u) &=&\frac{12\cosh(\alpha(\mu)\,u)}{1+2\mu^{2}}\exp\left[ -\frac{2\cosh(2\alpha(\mu)\,u)}{1+2\mu^{2}}\right] \nonumber \\ &+& 2\left(2\mu^{2}-1 -\sqrt{12-8\mu^{2}+(1+2\mu^{2})^{2}}\right)\nonumber \\ &\times& \frac{\cosh(3\alpha(\mu)\,u)}{1+2\mu^{2}}\exp\left[ -\frac{2\cosh(2\alpha(\mu)\,u)}{1+2\mu^{2}}\right], 
 \label{e53}\\
 \epsilon_{2}(\mu) &=& \frac{a(\mu)}{8\mu^{4}}(1 -2\mu^{2}) +\frac{11a(\mu)}{64\mu^{4}}(1+2\mu^{2})^{2} \nonumber \\ &+& \frac{a(\mu)}{16\mu^{4}}(1+2\mu^{2})\sqrt{(1+2\mu^{2})^{2} -8\mu^{2} +12}. 
 \label{e54}
 \end{eqnarray*}
 


\end{document}